\documentstyle[12pt,aasms4]{article}

\begin{document}

\def\etal{{\it et al.\/\ }}
\def\today{\ifcase\month
 &  &  &  &  &  \or January &  \or February \or March &  & \or April 
 &  &  &  &  &  \or May &  &  &  \or June &  &  \or July &  &  \or August
 &  &  &  &  &  \or September \or October & \or November \or December\fi
 &  &  &  &  &  \space\number\day, \number\year}
\def\ga{\lower 2pt \hbox{$\, \buildrel {\scriptstyle >}\over{\scriptstyle \sim}\,$}}
\def\la{\lower 2pt \hbox{$\, \buildrel {\scriptstyle <}\over{\scriptstyle \sim}\,$}}

\title{\bf REOBSERVATION OF CLOSE QSO GROUPS:
THE SIZE EVOLUTION AND SHAPE OF LYMAN ALPHA FOREST ABSORBERS$^1$}

\author{Arlin P. S. Crotts$^{2}$ and Yihu Fang}

\affil{Department of Astronomy, Columbia University, 538 W.~120th St., 
New York, NY~~10027}

\affil{\bigskip
$^1$ Based in part on observations made with the NASA/ESA {\it Hubble Space
Telescope}, obtained from the data archive at the Space Telescope Science
Institute, which is operated by AURA, Inc., under NASA contract NAS 5-26555.}

\affil{\bigskip
$^2$ Guest observer at the Kitt Peak National Observatory, National Optical
Astronomy Observatories, operated by AURA, Inc., for the NSF}

\authoremail{arlin@astro.columbia.edu, fang@astro.columbia.edu}

\begin{abstract}

In order to study the size and shape of the absorbers that lie in front of the
QSOs, in particular the Ly $\alpha$ forest, we present an analysis of 785
absorption lines in the spectra of five QSOs in close groupings: a pair
(LB9605: 1517+2357 at $z = 1.834$ and LB9612: 1517+2356 at $z = 1.903$, with a
separation of 102 arcsec between them) and a triplet (KP 76: 1623+2651A at $z =
2.467$, KP 77: 1623+2653 at $z = 2.526$, and KP 78: 1623+2651B at $z = 2.605$,
with separations of 127, 147 and 177 arcsec between pairs 76:78, 76:77 and
77:78, respectively).
Both of these QSO groups have been observed before, but these data represent a
drastic increase in signal-to-noise ratio and/or wavelength coverage over
earlier data, and provide a qualitatively different view of the nature of the
absorbers.
The pair samples a scale critical in determining the size upper bound of Ly
$\alpha$ absorbers, with significant leverage in redshift compared to previous
studies.
In the case of the triplet, this represents the spatially densest sample of Ly
$\alpha$ forest absorbers ever studied, and an almost ideally-suited probe of
the shape of absorbers.
We observe a significant number of Ly $\alpha$ lines in common between the
triplet sightlines, for lines stronger than rest equivalent width $W_o>
0.4$\AA\ (and no detected metal lines) and velocity differences up to
200~km~s$^{-1}$, corresponding to a two-point correlation function $\xi =
1.88^{+0.78}_{-0.50}$ on scales 0.5 to 0.8~$h^{-1}$~Mpc with $\langle z \rangle
=2.14$, and inconsistent at the 99.999\% level with the absence of any
clustering.
These data also show that a significant fraction of the $W_o > 0.4$\AA\ Ly
$\alpha$ forest absorbers span all three sightlines to the KP triplet,
indicating that the strong-lined absorbers are consistent with nearly round
shapes, chosen from a range of possible cylinders of different elongations.
This may be inconsistent with results from hydrodynamic/gravitational
simulations of H~I in the early Universe indicating that the theoretical
counterparts of Ly $\alpha$ forest clouds are long and filamentary.
Furthermore, there is a probable correlation of $W_o$ with $\Delta v$
suggestive of the clouds being flattened and expanding with the Hubble flow in
their long dimension, as would be indicative of sheets or filaments.
This is supported by the uniformity of linestrengths between the three
sightlines, for $W_o > 0.4$\AA.
We conclude, tentatively, that the $W_o > 0.4$\AA\ Ly $\alpha$ forest objects
are sheetlike.
In contrast, the weaker lines, 0.2\AA\ $> W_o > 0.4$\AA\, show no evidence of
spanning the sightlines of these groups, but have sizes significantly larger
than the luminous portions of galaxies, and C~IV absorbers as revealed by
closer-separation QSO pairs.
When the LB sightline pair is included with other pairs at different redshifts
and sightline separations, one finds no evidence for evolution of Ly $\alpha$
absorber size with redshift.
We also show that there is no evidence of large-scale structure in the Ly
$\alpha$ forest consistent with ionization of H~II by foreground QSOs as seen
in the spectrum of background QSOs (the ``foreground proximity effect'').
Finally, we see a marginal detection of the sightline two-point
cross-correlation function for C~IV lines $\xi = 2.05^{+1.82}_{-1.21}$ over
scales of 0.5 to 1~$h^{-1}$~Mpc.
This is significantly weaker than $\xi$ measured by auto-correlation along
single sightlines for 200~km~s$^{-1} < \Delta v < 600$~km~s$^{-1}$, suggesting
that most of the latter signal may be due to the internal motions within
absorbers which are smaller than 0.5~$h^{-1}$~Mpc.

\end{abstract}


\clearpage

\section{INTRODUCTION}

The past several years have seen rapid advances in our understanding of the
Ly $\alpha$ forest.
In part progress is based on the unprecedented spectroscopic capabilities of
the Hubble and Keck telescopes, but advances have also been made using 4-meter
class telescopes in measuring the size of the clouds.
The size and shape is crucial in establishing the spatial number density of the
cloud population, the ionization state of the clouds, their mass, and hence
their contribution to the mass density of the Universe.
Beyond more uncertain lensed QSO Ly $\alpha$ size limits (Foltz et al.~1984,
Smette et al.~1992, 1995), the first definite indication of large absorber
size came from observations of the 9.5 arcsec separation, $z=2.050$ QSO pair
1343+2640A/B (corresponding to a sightline separation of 39-40 $h^{-1}$~kpc proper distance, with $h = H_o/100$~km~s$^{-1}$~Mpc$^{-1}$ and $q_o=1/2$), indicating a absorber radius of $100-200~ h^{-1}$~kpc (Crotts et al.~1994,
hereafter Paper I; Bechtold et al.~1994, hereafter Paper II; Dinshaw et
al.~1994).
Subsequent observations of a lower redshift pair Q0107-0234/0107-0235 ($z =
0.952, 0.956$, separation$=301$ to $364~h^{-1}$~kpc proper distance)
suggested an even larger absirber size, and at lower redshift (Dinshaw et
al.~1995).
A treatment of previously published, higher redshift pairs (Shaver \& Robertson
1983, Crotts 1989, Elowitz, Green \& Impey 1995) at separations larger than
that of 1343+2640 revealed that absorbers must be either non-spherical,
clustered, or drawn from a distribution that is non-uniform in radius (Fang et
al.~1996 - Paper III), and also indicated the need for further data, either for
these pairs or new ones (yet to be discovered).
This approach is exploited in the current work, in \S 4.4.

The size of the absorbers implies that they contain a significant fraction of
the baryons in the Universe (Fang \& Crotts 1994, Rauch \& Haehnelt 1996, Paper
III).
As such they might be analyzed in detail using hydrodynamic/gravitational
simulations of the early Universe (Zhang, Anninos \& Norman~1995, Katz et
al.~1996, Miralda-Escud\'e et al.~1996).
Indeed, the size of the absorbers is key to identifying the corresponding
objects in the simulations e.g.~Cen et al.~1994, dealing with $\sim 20$~kpc
proper diameter clouds, in contrast to Miralda-Escud\'e et al.~1996, dealing
with clouds 100~kpc wide and 1~Mpc long.
One outcome of these simulations is also the possibility of predicting the
shape of the clouds, which can be compared to QSO triplet data to provide a
crucial test of the models, in \S 4.5.
Paper III also suggested that the shape of Ly $\alpha$ clouds might be 
studied directly using QSO triplets.
We present a measure of the absorber shape using triple sightlines in order to
facilitate this comparison.

Several theoretical papers have touched or even concentrated on structure on
small scales in the Ly $\alpha$ forest as measured by double sightlines.
Charlton, Churchill \& Linder et al.~(1995) suggest a number of tests sensitive
to the shape of Ly $\alpha$ clouds probed by pairs of sightlines.
These include a test of the correlation of the two neutral hydrogen column
densities ($N_{_{HI}}$) in the case where lines in different sightlines
correspond in redshift, a test based on $N_{_{HI}}$ of the detected line in the
case of an anticoincidence, where only one line is detected at a given
redshift, and a method for learning about the velocity field in the clouds by
measuring the velocity differences between sightlines.
These tests are constructed for comparison to idealized models of cloud shape
and kinematics.
Miralda-Escud\'e et al.~(1996) perform a detailed hydrodynamical/gravitational
numerical simulation of collapsing structure and gas reaction.
For sightline pairs passing through their model's volume, they calculate a
correlation coefficient describing the similarity of absorption between
sightlines as a function of transverse spatial separation and line-of-sight
velocity difference.
Cen and Simcoe (1997) investigate the shapes of clouds within simulations
like those of Miralda-Escud\'e et al., calculating the shapes of clouds
at different density contrast levels as well as their effective size.
They propose a test based on the correlation between velocity differences
between coincident lines versus the sightline separation for pairs.
They also present the spatial two-point correlation function for Ly $\alpha$
clouds of various densities (and presumably linestrengths) and study the
function's redshift evolution, and consider statistics based upon whether
absorption lines produced by sightlines passing through their model volume
arise in identical or different clouds (which is difficult to test
empirically in absorption spectra).
Charlton et al.~(1997) study model simulations (Zhang et al.~1995) analogous to
Miralda-Escud\'e et alia's and Cen and Simcoe's.
They compute size and shape measures analogous to those treated empirically in
Paper III, including the variation of inferred spherical (or disk-like) cloud
size as a function of pair sightline separation, and the related statistic of
line coincidence/anticoincidence ratio as a function of sightline separation.
They also recast the tests from Charlton et al.~(1995) in terms in this
simulation, as opposed to simplified cloud models.
We will apply several of the preceding tests in \S 4.8, as well as others that
we develop.

Multiple sightlines are also highly valuable in probing large scale structure,
in that sightlines separated by transverse distances smaller than the structure
should show large correlations with each other on these scales, since they
pierce the same features.
This remedies the problems of searching for voids with single sightlines
(Carswell \& Rees 1987, Crotts 1987, Duncan, Ostriker \& Bajtlik 1989, Rauch et
al.~1992), because multiple, well-sampled sightlines can provide a stronger
test for voids by placing more absorbers in a void-sized volume than could
possibly be obtained along a single sightline.
We study this in \S 4.3.
Furthermore, the effect of a foreground QSO upon the H~I distribution in front
of another QSO should be significant if the physics of the ``proximity effect''
(Bajtlik, Duncan \& Ostriker 1989) is as simple as supposed.
Some data exist on the foreground QSO proximity effect (Crotts 1989, Dobrzycki
\& Bechtold 1991, Fern\'andez-Soto et al.~1995), but they are inconclusive and
could easily be improved, as in \S 4.2.
Finally, large scale structure can also be sought in the metal-line
system distribution, and by cross-correlating multiple sightlines one
circumvents the possible ambiguity between internal velocity structure due to
motions within single absorber versus true spatial clustering of spatially
distinct objects.
These problems are addressed in \S 4.9.

\section{Observations}

Observations of all five QSOs were performed using the RC Spectrograph and T2KB
CCD on the Kitt Peak National Observatory's four-meter telescope on UT 1995
June 1-4, using the BL-420 and BL-450 gratings in second order for the
3170-4720\AA, 1.7\AA\ FWHM resolution setup and 4450-5750\AA, 1.4\AA\ FWHM
resolution setup, respectively.
Wavelengths are reduced to the vacuum heliocentric frame.
For the Q1623+2651A, 1623+2653, 1623+2651B data (KP 76, 77 and 78
respectively), spectrophotometric calibration was too uncertain to use.
{\it Hubble Space Telescope} observations of 1517+2356/1517+2357 (LB 9605 and
LB 9612 respectively) were made using the Faint Object Spectrograph G190H and
G270H setups as part of program GO 5320 of Foltz et al.~1994.
Data were also obtained on LB 9605 and 9612 UT 29-30 April 1992 by Elowitz et
al.~(1995) at the KPNO 4m/RC Spec at 3210-3590\AA, and they kindly let us coadd
their data with ours.
This involved recalibrating the wavelength scale, as there appeared large
shifts with respect to both the 1995 KPNO and {\it HST} wavelengths.
This was accomplished in four ways: by re-identifying lines in the Th-Ar
comparison spectrum accompanying the 1992 KPNO data, and by cross-correlating
the overlapping portions of the QSO spectra in the {\it HST} and 1995 KPNO
data with the 1992 KPNO data (and each other), as well as the night sky
spectra from the 1992 and 1995 KPNO data sets.
These show disagreements as large as 60~km~s$^{-1}$, so that over the region
3270-3340\AA\ the wavelength calibration might be uncertain by amounts as large
as this.
The resulting spectra, for KP 76, 77 and 78, are shown in Figures 1-3, and
for LB 9605/9612 in Figures 4-5.

\section{Analysis}

Continua were calculated and lines detected, de-blended and assigned
identifications according to Crotts (1989).
Deblending was performed using multiple gaussian fits instead of Voigt profiles
in nearly all cases, with the exception of resolved lines.
We use a lower than usual $S/N$ cutoff of $3.5 \sigma$ for line detections.
We are confident of this approach because we are able to check our results for
KP 77 against a high $S/N$ Keck HIRES spectrum of the object, the highest $S/N$
part of a larger sample collected for the triplet (Crotts, Burles \&
Tytler~1997).
At a $3.5 \sigma$ cutoff for the KPNO data, no false detections are found over
the 3900-5700\AA\ overlap between the two data sets.
In fact the KPNO 4m observations do a good job of detecting all obvious lines
in the Keck spectrum (except for some very weak lines in the red KPNO 4m
spectrum past 4700\AA), while, of course, not resolving very close lines.
We are fairly confident of our linelists, therefore, and expect approximately
four lines out of our 785 to be false detections due to statistical
fluctuations.

\subsection{Absorption Line Identification}

We try to be complete as possible in identifying metal-line systems, since
stray metal lines might contaminate the statistical properties of Ly $\alpha$
samples, particularly if metal-line systems are redshift correlated.
We follow the procedure of Crotts (1989) and list all ``definite'',
``probable'' and ``possible'' systems (the later denoted by ``?'') thereby
dividing the systems into classes by probability of being reproduced in a
random linelist having the same global distribution as real QSO absorption
lines.

We identify the following metal-line and Lyman series absorption line systems,
detailed in the linelists (Tables 1-5) and listed by QSO below.

\noindent {\it KP 76:}

This QSO is surprisingly lacking in absorption lines and systems.
Only two well-established metal-containing systems are seen:

\noindent $z_{ab} = 2.11226$.--
This system shows strong Ly $\alpha$, C~IV and Si~III, as well as a firm Si~IV
doublet, plus Si~II $\lambda$1260 and C~II $\lambda$1334.
Four of these eight lines occur beyond the Ly $\alpha$ forest.

\noindent $z_{ab} = 2.24563$.--
Very strong Ly $\alpha$ and Ly $\beta$ compose this system, along with weak
C~IV $\lambda$1548, Si~III and C~I $\lambda$1277.
Ly $\alpha$ is significantly offset from the other lines, suggesting that it is
contaminated by another line.

\noindent $z_{ab} = 1.93778$?, 2.40484? and 2.44125?.--
KP 76 shows two Ly $\alpha$/Ly $\beta$ pairs at $z = 2.40484$ and 2.44125, both
with much weaker associated Si~III $\lambda$1206, but no detected C~IV.
There is a similar system at $z = 1.93778$, but Ly $\beta$ is below the short
wavelength limit.

\bigskip

\noindent {\it KP 77:}

\noindent $z_{ab} = 0.88720$.--
This is a strong system marked by many lines redward of the Ly $\alpha$ forest.
It is peculiar that the Mn~II lines are offset several hundred km/s from the
five Fe~II and four Mg~II and Mg~I lines.
The Mg~II doublet is mixed into a complex of structure near 5285\AA, but is the
strongest contribution to this complex.

\noindent $z_{ab} = 2.40060$.--
Ly $\alpha$ is strong, as is C~IV in this system, although the 1550\AA\ line is
confused with the Mg~II doublet at 0.88720.
Weaker Ly $\beta$, Si~III $\lambda$1206 and a possible O~VI doublet also
support the system.

\noindent $z_{ab} = 2.40602$?--
Further structure in the complex at 5285\AA\ is explained by a C~IV doublet at
$z_{ab} \approx 2.406$.
A strong Ly $\alpha$ line sits at slightly higher redshift.

The exact redshift of this system will benefit from higher resolution data on
the complex mixing this and the last two systems.

\noindent $z_{ab} = 1.87952$, 1.97331, 2.05050, 2.05380, 2.16128, 2.24455 and
2.52900.--
These are all examples of C~IV doublets appearing redward of the Ly $\alpha$
emission line, accompanied by strong Ly $\alpha$ at the same redshift.
The $z_{ab} = 1.97331$ system also shows possible Si~III $\lambda$1206.
The $z_{ab} = 2.053$ C~IV doublet's redshift sits on the blue wing of the
corresponding Ly $\alpha$ line.
The C~IV doublet at $z_{ab} = 2.1616$ is blended with other lines, hence less
certain.
The $z_{ab} = 2.52900$ system is at high enough redshift so that Ly $\beta$ is
also detected, and possible Fe lines.

\noindent $z_{ab} = 2.44490$ and 1.67130?--
These are Lyman series systems with weak C~IV $\lambda$1548.
The $z_{ab} = 2.44490$ system is more definite having Ly $\beta$ and possible
Ly $\gamma$ as well as Ly $\alpha$ (the only Lyman series line at $z_{ab} = 
1.67130$ appearing above the short wavelength cutoff).

\noindent $z_{ab} = 2.40966$?, 2.46345? and 2.47424?--
These are strong Ly $\alpha$/Ly $\beta$ pairs, without other supporting lines.

\bigskip

\noindent {\it KP 78}:

\noindent $z_{ab} = 2.09428$.--
This is a very strong, probably damped, Ly $\alpha$ absorber with strong
detections of the C~IV and Si~IV doublets, Si~III $\lambda$1206, C~II
$\lambda$1334, weaker Al~II $\lambda$1670 and probable contributions from
Si~II $\lambda$1260 and Fe~II $\lambda$1122.
The Ly $\alpha$ centroid is offset about 80~km/s from the other lines.

\noindent $z_{ab} = 2.23925$.--
This system is defined by strong C~IV and Si~IV doublets as well as C~II
$\lambda$1334, plus possible contributions from C~I $\lambda$1277 and Fe~III
$\lambda$1122.
It shares a strong Ly $\alpha$ line with the $z_{ab} = 2.24173$ below.

\noindent $z_{ab} = 2.24173$.--
As well as very strong Ly $\alpha$ and C~IV, this system shows C~I
$\lambda$1277, C~II $\lambda$1334, the N~V doublet, Si~IV $\lambda$1393, Si~III
$\lambda$1206, the strongest lines of Si~II (1260\AA, 1304\AA\ and 1190\AA),
plus possible Fe~III $\lambda$1122.

\noindent $z_{ab} = 2.55106$.--
This system shows strong Ly $\alpha$, $\beta$ and $\gamma$, Si~III and weak
C~IV $\lambda$1548, Si~IV $\lambda\lambda$1393, 1402, Si~II (1260\AA, 1304\AA\
and 1190\AA) and possible C~II $\lambda\lambda$1036, 1334, C~III $\lambda$977
and Fe~III $\lambda$1122.

\noindent $z_{ab} = 2.57482$.--
There is no C~IV detected but moderate Ly $\alpha$, $\beta$, $\gamma$ and
$\delta$.
Also possible are Si~III, Si~IV $\lambda$1393, Fe~III $\lambda$1122 and Fe~II
$\lambda$1144.

\noindent $z_{ab} = 1.98490$, 2.04240 and 2.09592.--
These are all C~IV doublets outside of the Ly $\alpha$ forest, plus strong Ly
$\alpha$ lines.
The $z = 1.98490$ system also contains Ly $\beta$.
The redshift 2.09592 system shares Ly $\alpha$ with 2.09428 above and probably
also shows Si~III.

\noindent $z_{ab} = 1.93770$? and 2.06117?--
These are possible systems consisting of weak C~IV $\lambda$1548 outside of the
forest plus strong Ly $\alpha$.

\noindent $z_{ab} =
2.36526$?, 2.42738?, 2.44138?, 2.45570?, 2.45801?, 2.53649?, 2.53941,
2.54356?, 2.54879?, 2.56715 and 2.59458?--
These are ``possible'' and probable Ly $\alpha$, $\beta$ pairs.
The $z_{ab} = 2.53941$ and 2.56715 also show Ly $\gamma$.
The 2.59458 system is only suspected.
If this line is Ly $\alpha$, one should expect to see Ly $\beta$ at 3687.04\AA.
There is a marginally detected line there, at about $W_{obs} = 0.3$\AA, so this
is plausibly consistent with 4369.82\AA\ being Ly $\alpha$.

\bigskip

\noindent {\it LB 9605}:

\noindent $z_{ab} = -0.00055$.--
This Galactic system is marked by the Mg~II doublet, Ca~II $\lambda$3934,
Mg~I $\lambda$2852, Al~II $\lambda$1670, the four strongest lines of Fe~II,
two of Fe~I and three of Mn~II.
Note that a similar system is seen in LB~9612.

\noindent $z_{ab} = 0.73825$.--
This is composed of Ly $\alpha$, $\beta$ and $\gamma$, plus the C~IV doublet
and Si~IV $\lambda$1393.
All of these lines fall in the forest but appear to be unambiguous.

\noindent $z_{ab} = 1.02780$.--
This shows Ly $\alpha$, $\beta$, $\gamma$ and $\delta$, the C~IV doublet, Si~IV
$\lambda$1393, Si~II $\lambda$1206, C~III $\lambda$977 and a likely O~VI
doublet.
Most of these lines are blended with components of other systems, and none land
beyond the forest.
(None would be expected.)

\noindent $z_{ab} = 1.51350$.--
A weak C~IV doublet lands outside the forest, while the other lines consist of
Ly $\alpha$, probable Ly $\beta$, Si~II $\lambda$1260, C~I $\lambda$1277, and
possible Fe~II $\lambda$1144, Fe~III $\lambda$1122 and the O~VI doublet.

\noindent $z_{ab} = 1.59590$.--
This is a strong system consisting of the first ten Lyman series lines plus
C~IV $\lambda$1548.
The absence of a Ly limit feature ($\tau < 0.1$) implies
$N_{_{HI}} \la 1.5 \times 10^{16} cm^{-2}$.

\noindent $z_{ab} = 0.41063$.--
This is a probable C~IV doublet in the forest associated with a strong Ly
$\alpha$ line.

\noindent $z_{ab} = 1.19268$.--
This is a probable Ly $\alpha$/Ly $\beta$ pair with a C~IV $\lambda$1548 line
beyond the forest.

\noindent $z_{ab} = 1.80465$.--
Another strong system of the first seven Lyman series lines, it shows no metal
lines.

\noindent $z_{ab} = 1.01602$, 1.07973, 1.30150, 1.60295, 1.68573,
1.72400?, 1.84560.--
These consist of at least the first three Lyman series lines in a pure hydrogen
systems.
Additionally, one finds Ly $\delta$ for 1.30150, 1.60295, 1.07973 and 1.84560,
and Ly $\epsilon$ for 1.07973.
There is a possible C~IV $\lambda$1550 line at $z_{ab} = 1.30142$.
The $z_{ab} = 1.72400$ system is more uncertainty due to ambiguity in some of
its line identifications.

\noindent $z_{ab} = 0.70010$?, 0.96730?, 1.16401?, 1.19274?, 1.42858?,
1.55105?, 1.58226?, 1.62531?,\\
\noindent 1.68137?, 1.70001?, 1.73948?, 1.74690?, 1.75165?,
1.77245?, 1.78948?, 1.82251?--
These are possible Ly $\alpha$/Ly $\beta$ pairs; all have strong Ly $\alpha$.
The 1.82251 and 0.96730 systems also show Si~III $\lambda$1206.

\bigskip

\noindent {\it LB 9612}:

\noindent $z_{ab} = -0.00040$.--
Like LB 9605, this spectrum shows a strong Galactic system.
It includes the Ca~II and Mg~II doublets, Mg I $\lambda$2853, Ca~I
$\lambda$2722, Fe~I $\lambda$2523 and three strong lines of Fe~II (2382\AA,
2586\AA\ and 2600\AA).
Fe~I $\lambda$2484 is lost in the Ly break at 2484\AA.

\noindent $z_{ab} = 0.25155$?--
Another possible low-redshift system consists of the Mg~II doublet and Mg~I
$\lambda$2852, all in the forest.

\noindent $z_{ab} = 0.73690$.--
This consists of Ly $\alpha$, $\beta$, $\gamma$ and a C~IV doublet, all in the
forest.

\noindent $z_{ab} = 1.05998$.--
This system contains C~IV and Si~IV doublets, Ly $\alpha$ and $\beta$,
and possible C~I $\lambda$1656.

\noindent $z_{ab} = 1.12625$.--
This shows Ly $\alpha$, a C~IV doublet, Si~II $\lambda$1393, C~I $\lambda$1656
and possible N~V $\lambda$1238, Si~III $\lambda$1206 and C~I $\lambda$1277.

\noindent $z_{ab} = 1.30090$.--
This probable system contains Ly $\alpha$, $\beta$ and C~III $\lambda$977
confused with other systems, Si~II $\lambda$1260 and, outside the forest, C~IV
$\lambda$1548.

\noindent $z_{ab} = 1.41451$.--
This is marked by strong Ly $\alpha$, a C~IV doublet (outside the forest) and a
Si~IV doublet.

\noindent $z_{ab} = 1.42671$.--
This consists of a strong Lyman series to Ly 8, plus C~IV $\lambda$1548.
The system's Ly $\beta$ occurs at the Ly break at 2484\AA.
Ly 6 at 2258\AA\ is marginally detected.

\noindent $z_{ab} = 1.55414$.--
This includes Ly $\alpha$, Ly $\beta$, a C~IV doublet and C~II $\lambda$1334.

\noindent $z_{ab} = 1.72398$.--
This is a very strong system with Lyman series lines up to at least Ly 10, and
probably includes a blend of higher terms in the series.
These are associated with a $N_{_{HI}} \approx 4\times
10^{17}$~cm$^{-2}$ determined from the Ly limit drop.
Metal lines include C~IV $\lambda$1548, C~III $\lambda$977 and C~I lines
(1277\AA\ and 1656\AA).

\noindent $z_{ab} = 1.88690$.--
In addition to a Ly series extending to Ly 10 and a Ly limit break
corresponding to $N_{_{HI}} \approx 1 \times 10^{17}$~cm$^{-2}$, this system
shows Si~III $\lambda$1206, C~IV $\lambda$1548 and possible C~III $\lambda$977
and O~VI $\lambda$1031.

\noindent $z_{ab} = 1.75930$ and 1.79961.--
These are two Ly series systems extending to Ly 7 and Ly $\epsilon$,
respectively.

\noindent $z_{ab} = 1.43510$?, 1.43964?, 1.62398?, 1.64890?, 1.70809? and
1.71516?--
These are possible Ly $\alpha$/Ly $\beta$ pairs.

\bigskip

Both LB 9605 and LB 9612 have significantly more identified redshift systems
per unit $z$ than KP 76, 77 and 78, probably due to the greater wavelength
coverage of these two spectra.
Many of these systems would be unlikely to be recognized in the KP spectra.
The consequences of this will be discussed in the next subsection.

\subsection{Lyman Alpha Line Sample}

Despite the large number of lines detected at wavelengths shorter than Ly
$\alpha$ emission in LB 9605 and 9612, relatively few are due to unadulterated
Ly $\alpha$ absorption, at least below certain wavelengths.
This is particularly true for LB 9612, which suffers a nearly complete loss of
flux below 2490\AA\ due to one Ly limit system, and a significant drop below
2670\AA\ due to another.
Equally serious, however, is the contamination of a large stretch of spectrum
by higher Ly series lines and metal lines, many associated with these two Ly
limits.
Of the 45 lines between 2490\AA\ and 2900\AA, only six are explained by
uncontaminated Ly $\alpha$ lines, whereas above 2900\AA\ only 29\% of the lines
in the forest can be explained (even in part) by lines other than Ly $\alpha$.
Similar behavior, although not so drastic given the absence of Ly limit
systems, is seen in LB 9605.
This implies that the only useful sample of Ly $\alpha$ lines for comparing
the Ly $\alpha$ distributions in LB 9605 and 9612 is in the wavelength range
2900-3445\AA\ ($z = 1.39$ to 1.83, with $\langle z \rangle = 1.62$), roughly
the range between Ly $\beta$ and Ly $\alpha$ emission.
(In practice, we use observed wavelength 2900\AA\ to the wavelength 1220\AA\ in
the reference frame of the QSO, accounting for infall towards the QSO by up to
1000~km~s$^{-1}$.)

We apply a similar constraint to the KP 76, 77, 78 triplet.
In practice we consider those lines with wavelengths of 1020\AA\ to 1220\AA\ in
the reference frame of the QSO, unless explicitly stated otherwise.
While their spectra below Ly $\beta$ emission are not so obviously contaminated
by non-Ly $\alpha$ lines, this might be due to our ignorance of further
metal-line systems because of the smaller wavelength coverage in these spectra
compared to the LB pair.
The Ly $\beta$ to $\alpha$ emission line range from above constrains these
samples such that all three overlap for redshifts 2.02 to 2.48.

For lines at redshifts lower than $z=2.02$, the sensitivity of our data is also
declining, but for some purposes, sensitivity cutoffs as large as $W_0=
0.4$\AA\ are useful.
Also, for some purposes, we might be less worried about metal-line
contamination, especially since we show that the metal-line system redshifts
are weakly correlated between sightlines.
In a limited number of specified cases, we impose a cutoff at $\lambda =
3350$\AA\ (or $z=1.756$), below which the sensitivity in KP 78 drops below
$W_0 = 0.4$\AA.
Given a line-of-sight number density evolution $N(z) \propto (1+z)^\gamma$,
with an assumed $\gamma = 2.1$ (see \S 4.3), we study an average $\langle z
\rangle = 2.25$ for the generally-used restrictive sample, and $\langle z
\rangle = 2.14$ for the less-used, larger redshift range. 

We take as Ly $\alpha$ lines all those between the wavelengths listed above and
not otherwise identified as a metal line (although they can be the Ly $\alpha$
component of a metal-line redshift system).
This is the ``pure'' Ly $\alpha$ sample.
The ``contaminated'' sample is one in which a detectable contribution is
suspected from a metal line from another redshift system,
but the presence of a Ly $\alpha$ line is inferred
from the strength of the actual line above that of the inferred metal line.

For unresolved Ly $\alpha$ lines in the KP triplet sample, then, the
completeness cutoff (the $3.5\sigma$ threshold plus another $2 \sigma$ to
assure completeness) in each 1020-1220\AA\ region is about $W_o = 0.19$, 0.09
and 0.15\AA, respectively.
For LB 9605 and 9612, the sample reaches $W_o \approx 0.12$ and 0.13\AA,
respectively (except for a small interval $1.691 < z < 1.768$ for LB 9605,
where the threshold is as high as 0.4\AA), with the same caveats regarding
uncrowded and unresolved lines.
In some of our treatment below, we will discuss thresholds as low as $W_o =
0.1$\AA, knowing that this falls slightly short of our completeness condition,
but most interesting results apply to cutoffs of $W_o = 0.2$\AA, 0.4\AA, or
higher.
The sensitivity of the spectra and our various Ly $\alpha$
samples are further described in Table 6.

\section{Results}

These higher quality data allow us to improve several unique measurements made
in Crotts (1989) using the KP triplet, plus several new tests that we
apply for the first time.
Sections 4.1 through 4.3 deal with tests originally developed for the triplet
in Crotts (1989) and section 4.4 applies cloud size techniques developed in
Papers II and III to the LB pair for the first time, as well as the improved
triplet data set.
For this reason we reserve detailed discussion of techniques for section 4.5
and later, and refer the reader to these previous papers for earlier
developments.

\subsection{Ly $\alpha$ Velocity Cross Correlation}

Following Crotts (1989), we can compute the spatial two-point function of Ly
$\alpha$ absorbers by cross-correlating in velocity the distribution of Ly
$\alpha$ lines (``pure'' and ``contaminated'').
The resulting pairs are binned at 50 km s$^{-1}$ intervals with cutoffs of,
alternately, $W_o > 0.1$, 0.2, 0.4 and 0.8\AA.
For the triplet (all three sightline pairs summed together) and the pair, the
resulting cross-correlation pair count as a function of velocity difference
$\Delta v$ is shown in Figures 6a and 6b, respectively.
For the triplet, the $\langle z \rangle = 2.25$ sample is shown.

In Figure 6a for the triplet, there is no apparent structure for weaker lines
e.g.~samples with cutoffs at $W_o = 0.1$\AA\ or 0.2\AA, but structure is
apparent for lines with $W_o > 0.4$\AA\ or 0.8\AA.
For $W_o > 0.4$\AA, there are an average of 2.97 pairs per 50 km s$^{-1}$ bin
in the first 3000 km s$^{-1}$ of $\Delta v$ (and 2.83 per bin over
$3000~$km s$^{-1} < \Delta v < 6000~$km s$^{-1}$), so that the appearance of
eight pairs (or more) in the first 50 km s$^{-1}$ bin as a random fluctuation
is ruled out at the 99.0\% level.
(This is the Poisson confidence level for excluding the hypothesis that there
is no intrinsic excess in the first bin, as might be substantiated by larger
samples.)~
There are also 20 pairs in the first 200 km s$^{-1}$, which is ruled out as a
random fluctuation at the 98.4\% level.
Thus there is evidence for a cross-correlation signal in the smallest velocity
bins for pairs of $W_o > 0.4$\AA\ lines, at about the $2.5 \sigma$ level.
This can be described by a two-point correlation function, averaged over
proper separations of 496 to 720 $h^{-1}$kpc and with $\langle z \rangle =
2.25$ of $\xi = 0.72_{-0.38}^{+0.48}$ (68\% confidence limits - {\it not}
directly translatable into gaussian standard deviations for such a small
sample).
For the extended, $\langle z \rangle = 2.14$ sample, the clustering signal is
stronger, both in magnitude and statistical significance: 29 pairs in the first
200~km~s$^{-1}$ bin versus 15.6 expected, implying $\xi =0.86^{+0.41}_{-0.28}$,
inconsistent with $\xi = 0$ at the 99.92\% level.
When only ``pure'' Ly $\alpha$ lines are considered, there are 26 in the first
bin with 11.7 expected, implying $\xi = 1.23^{+0.53}_{-0.36}$, inconsistent
with zero at the 99.991\% level.

How important is the elimination of all Ly $\alpha$ lines with associated
metal lines at the same $z$?
We did this for the extended, $\langle z \rangle = 2.14$ sample and found 21
pairs in the first 200~km~s$^{-1}$ bin versus 7.3 expected, giving $\xi =
1.88^{+0.78}_{-0.50}$, which is inconsistent with the no-clustering hypothesis
at the 99.999\% level.
Hence, the exclusion of the metal-containing systems makes the signal
marginally stronger.

There is a weak signal for larger $\Delta v$ and stronger lines.
Nine $W_o > 0.8$\AA\ pairs land within $\Delta v < 600~$km s$^{-1}$ versus the
mean expectation of only 3.6, a result expected in only 1.2\% of random cases.
Both of these results, for $W_o > 0.8$\AA\ and 0.4\AA\ lines, are very similar
to results found by Crotts (1989) for a significantly smaller sample.

For the LB pair, the results are less impressive.
For $W_o > 0.4$\AA, there are no pairs in the first 50 km s$^{-1}$ bin, and
only three pairs in the first 200 km s$^{-1}$ versus a mean expectation of
1.9.   
In the first 300 km s$^{-1}$, there are six pairs versus 2.8 expected, which is
ruled out as being random at only the 93.5\% level.
There is a larger excess in the $W_o > 0.2$\AA\ sample, nine pairs versus 6.0
mean expectation in the first 200 km s$^{-1}$, but it is even less
statistically significant.  

We conclude that the $\Delta v < 150-200$ km s$^{-1}$ cross-correlation signal
obvious in Paper III, Figure 4 for QSO pairs closer than $400~h^{-1}$ kpc
persists to over $700~h^{-1}$ kpc in the KP triplet, despite the weakness of
the signal seen in the LB pair at separation $S \approx 430~h^{-1}$ kpc.

\subsection{Large Scale Structure in the Ly $\alpha$ Distribution}

As discussed in Crotts (1989), the distributions of the three Ly $\alpha$
forests in the triplet can be combined into a single probe of structures much
larger than their sightline separations of 0.5-0.7 $h^{-1}$ Mpc.
These are constructed by running bins of different widths along complete Ly
$\alpha$ line lists, counting the number of lines in each bin.
The results are shown in Figures 7a and 7b for the $W_o > 0.4$\AA\ and
$W_o > 0.1$\AA\ samples, respectively.
The bin width alternates between 15, 30 and 45 $h^{-1}$ Mpc (for $q_o = 1/2$)
and the bin center is stepped in redshift every 1/4 of the bin width.
In Figure 7a, two prominent underdensities occur, one at $z = 2.08$ and another
at $z = 2.37$; from the bin width plot in which they are most prominent, the
appear to have widths of about 15 and 30 $h^{-1}$ Mpc, respectively.
How statistically significant are they?
The first dips to zero counts when the mean is 4.5 and represents one bin
among 15 independent bins across the total redshift range, leading to an {\it
a priori} probability of finding a void this marked at about the 17\% level due
to random fluctuation.
The $z = 2.37$ feature drops to a count of three versus an average of eight in
one 30 $h^{-1}$ Mpc bin (versus seven independent ones), for an {\it a priori}
probability of 30\%.

Note that these features are of roughly the same size as underdensities seen
toward other QSOs: 0420-388 (Crotts 1987, Rauch et al.~1992) and 0302-0019
(Dobrzycki \& Bechtold 1991).
Note that the same sort of plot for $W_o > 0.1$\AA\ line (Fig. 7b) shows no new
features and the previously mentioned underdensities are washed out.
If such structures are real, they are traced more by the stronger lines, in a
way similar to the small scale structure seen in \S 4.1.

Neither of these features is significant enough to stand by themselves as a
detection.
It is interesting nonetheless to ask if they are associated with foreground
QSOs, as might be expected if several bright QSOs sit in the foreground of
this triplet and thereby destroy the neutral hydrogen at their redshift, or,
if voids might exist as in the galaxy distribution, bounded by walls of more
condensed objects, such as QSOs, perhaps.
A search of recent QSO catalogs (e.g.~Hewitt \& Burbidge 1993) reveals one
faint, possible QSO (KP 70) at $z = 2.1$(?) and another at $z =2.4$(?) (KP 73).
Both have $V \ga 20$ and sit too far from the triplet in the sky (angular
distances corresponding to about $6 ~ h^{-1}$ Mpc and $3 ~ h^{-1}$ Mpc proper
separation, respectively) to be likely causes for such large voids (unless the
flux we see is not representative of the flux experienced by observers in other
directions, either due to variability of the QSOs or anisotropic radiation
patterns).
These do not seem likely to produce such underdensities due to a foreground
QSO proximity effect; for this reason we are searching for other QSOs in the
field (Crotts 1998).

\subsection{Foreground QSO Proximity Effect}

A more direct approach (Crotts 1989, Bajtlik et al.~1988 with correction found
in Crotts 1989) to estimating the effects of QSOs on the absorbers along the
sightlines to background sources is to compute the radiation field from the
known, bright QSOs in the foreground and their effects on neutral hydrogen in
their vicinity.
This can be compared to the actual number density of lines seen towards the
three background QSOs in the triplet, and the model can thus be tested.
Over the relevant redshift range of interest (see Table 7), all QSO spectra
are sensitive to $W_o = 0.1$\AA.
These foreground QSOs include the triplet and the $z = 2.183, ~ V = 19.6$ QSO
KP 79, sitting about $1.2 ~ h^{-1}$ Mpc in proper distance to one side
(closest to KP 77 and 78).
We make the assumption that the Ly $\alpha$ clouds are distributed uniformly
except for the general evolution of line-of-sight number density with redshift,
$n (z) = N_* ( 1 + z)^\gamma$, where only lines with $W_o > 0.1$\AA\ are
counted, $\gamma = 2.1$ (intermediate between two recent determinations:
Bechtold 1994, Lu, Wolfe \& Turnshek 1991), and $N_*$ is adjusted to maintain
equal total lines in the model and triplet sample (for $1.99 < z < 2.49$).
(The measured number lands within 10\% of predictions from the literature,
after adjusting for different sample sensitivities.)
We ignore momentarily the possibility that the proximity effect is modified by
large scale structure influencing the local number density of Ly $\alpha$ lines
(Loeb \& Eisenstein 1995).

To illustrate the correlation of observed $n (z)$ with predicted deficits of
lines due to the proximity effect, we present Figure 8, which shows the general
evolution of $n_\gamma(z)$ (dotted line) absent the effects of local radiation,
the altered $n_p (z)$ (solid curve) predicted by the proximity model of Bajtlik
et al.~(1988) assuming $J_{21} = 1$, and the actually observed density of lines
$n_o$ in redshift bins selected to be equally spaced in redshift but
well-placed with respect to foreground QSOs (crosses showing $z$ intervals and
1$\sigma$ error bars in $n_o$).
Figure 9 shows the same information for LB 9612, which has LB 9605 in the
foreground, as well as a minor contribution from the $V=18.2$, $z=1.818$
LB~9615 sitting 9 $h^{-1}$ Mpc to one side.
Note that there is some correlation with the model $n$ in the LB pair, but in
most cases the counts and model are {\it anti-correlated} for the triplet.
The signal involved is still small compared to the errors, so we weight the
data in a more optimal way, first explained in Crotts (1989), and repeated
below.

The weighting factor, applied to each line observed in the redshift interval
with significant $n_\gamma-n_p$, is just the line density deficit $w = 1 -
n_p/n_\gamma$ at that redshift ({\it not} $\omega$ from Bajtlik et al.~[1988]).
The sum over observed lines is compared to predictions from the two models:
$w$ integrated over $n_\gamma(z)$, and $w$ integrated over $n_p (z)$.
Lines are only considered and integrals only calculated over regions where
$w > 0.1$.
The results are shown in Table 7.

As in Crotts (1989), the observed signal is consistent with the no-proximity
model, and lends no support to the model including foreground ionization, being
discrepant with that model at the $2.4\sigma$ level.
This result persists despite the new spectroscopy, inclusion of lines with
0.1\AA$ < W_o < 0.2$\AA, and the addition of LB 9612.
Two other papers, Dobrzycki \& Bechtold (1991) and Fern\'andez-Soto et
al.~(1995) bear on the foreground-QSO proximity effect.
The latter detects a marginally significant signal consistent with a foreground
effect, while the phenomenon seen by Dobrzycki \& Bechtold is much too strong
to be explained by a simple interpretation of QSO ionization effects for the
brightness seen for the foreground QSO.
As Loeb \& Eisenstein (1995) point out, the proximity effect in the case of a
single QSO can be altered by the effects of large scale structure in the
immediate vicinity of the QSO, in the sense that cluster produces more Ly
$\alpha$ lines near the QSO's Ly $\alpha$ emission.
This effect is strongest for faint QSOs, where ionization is weak.

Even though the QSOs studied here tend to be faint because of the requirement
that they reside in close pairs, the Loeb \& Eisenstein result does not explain
why these might be so discrepant with the Bajtlik et al.~model while bright
QSOs are not.
Furthermore, we have chosen a value of $J_{21}$ that is often regarded as low.
One should recall that anisotropic radiation by the QSO or long term
variability on timescales of about $10^5$~y (Crotts 1989) are
potential means by which the foreground QSO proximity effect can produce a
signal that is out of proportion with the observed flux, while the direct
(single QSO) proximity effect involves Ly $\alpha$ clouds along the observed
sightline to the ionizing source and photons emitted at the same time as the
ionizing photons.
If these factors are manifest here, a much larger sample will be needed to
reveal them.
The current sample does not lend additional support to the ionization
interpretation of the proximity effect by way of the foreground QSO test.

\subsection{Ly $\alpha$ Absorber Size Estimates}

The power of QSO pairs in providing transverse information is crucial to
finding the size of absorption clouds, particularly in the forest.
We (Papers II and III) have constructed a simple statistical measure of cloud
size based on the working assumption that Ly $\alpha$ are unclustered spheres
of uniform radius.
In Paper III, however, we show that this assumption cannot be
completely accurate
because the inferred cloud radius $R$ is a function of the QSO pair separation
$S$, contrary to the basic assumption.
Much of this failure is based on the behavior of absorbers in the KP QSO
triplet, which we re-examine here.
Furthermore, Dinshaw et al.~1995 also consider the pair Q0107-0234/0107-0235,
which is of much lower redshift than the other pairs that have been studied,
and suggest that the cloud size increases with lower $z$.
The LB pair allows us to test this possibility.

Our technique consists of an analysis of ``hit'' statistics, a hit consisting
of a line above a set $W_o$ threshold detected in both sightlines of a QSO
pair, with a velocity difference between the two absorption line redshifts
less than a velocity difference cutoff.
Figure 4 of Paper III shows that a cutoff of $\Delta v = 150-200$~km~s$^{-1}$
is strongly suggested by the the presence of a strong cross-correlational
signal between all published sightlines up to the scale of the KP triplet's
separations, and is further borne out in \S 4.3 by the clustering feature at
$\Delta v < 200$~km~s$^{-1}$ seen in the triplet.

When a line is seen in one QSO spectrum, but no line above the $W_o$ threshold
is seen within 200~km~s$^{-1}$ in the other (and this can be established with
greater than $3.5\sigma$ certainty) it is registered as a ``miss.''
(If such a situation is not established with $3.5\sigma$ certainty, it is
``null.'')
We assume in turn one of three cosmological models: $(\Omega_0, \Lambda_0 =
\Lambda/3H^2) = (1,~0)$, (0.1, 0) and (0.1, 0.9).
In all three cases, for the relevant ranges of redshifts, we can consider the
separation between sightlines to be nearly constant, with an order-unity
ratio between sightline separations (listed in Fang et al.~1996 for all QSO
pairs considered here except the LB pair) for the three cosmological cases.
For the LB pair. multiply the (1,0) value by 1.38 for (0.1,0) and 1.9 for
(0.1, 0.9).

We limit our sample to lines with $W_o \ge 0.4$\AA\ for all QSOs, and include
the ``contaminated'' Ly $\alpha$ lines for the triplet and LB pair.
In Table 8, we show the Ly$\alpha$ forest redshift ranges, angular separations,
proper separation range for (1,~0), hit and miss counts, inferred 95\%
confidence intervals, and median predicted cloud radii ($q_o=1/2$) for the QSO
pairs Q1343+2640 (Papers II and III), Q0307-1931/0307-1932
(Shaver \& Robertson 1983), Q0107-0234/0107-0235 (Dinshaw et al.~1995), and new
values for Q1623+2651A/1623+2653/1623+2651B, Q1517+2357/1517+2356 and
Q1026-0045A/B.
(See note added in proof.)

[Note added in proof: Q1026-0045A/B (Petitjean et al.~1998), like
Q1517+2357/1517+2356
are two low-redshift QSOs in a close pair observed by the FOS on $HST$ using
the G270H grating.  These data are now included in Table 8, Figure 10 and the
results of this section.  We have reanalyzed the linelist and spectra of this
pair, imposing the same $W_0 = 0.4$\AA\ cutoff as for the other pairs, which
coincidentally results in a somewhat smaller $R$ value than do
the $W_0$ cutoffs used by Petitjean et al.  Nonetheless, adopting their values
does not change the results of this section significantly.]

The estimates for the KP triplet were computed by taking each pair of QSOs
separately; strictly speaking they are not quite independent.
Also, for these samples there are significant numbers of accidental hits; these
are corrected as follows: we assume a Poisson distribution for the number of
random hits, with the mean of $N_{rand}=11.6/3=3.87$ random hits per QSO pair.
The number of ``real'' (non-random) hits, $N_{real}$ are given by the observed
number minus the random component.
Since the real component must be non-negative, that part of the distribution
with $N_{rand}$ greater than $N_h$ is included in the $N_{real}=0$ bin.
Each $N_{real}>0$ produces its own probability density distribution in
$\cal P$$(R)$ (as in Paper III) while cases where $N_{real}\le 0$ produce only
an upper limit in $R$.
(The probability distributions for cases where $N_{real}\le 0$ are taken as
constant in $R$: $\cal P$$(R) = \lbrace^{~0,~ R < S/2}_{~constant,~R\ge S/2}$.
Fortunately, these cases are a small fraction of the total.)~
The median $R$ value and corresponding $R$ confidence intervals are computed
by taking an average of the probability distributions corresponding to a
different $N_{real}$, weighted by this truncated Poisson distribution.
The results are slightly smaller in median $R$, and with smaller errors, than
those derived from Crotts (1989) data in Paper III.

The LB pair is particularly interesting because it lands mid-range in the span
of $S$ values from pre-existing pair observations, but is at significantly
lower redshift than average.
Unlike the lower redshift pair Q0107-0234/0107-0235 (Dinshaw et al.~1995),
however, it does not imply $R$ values significantly higher for pairs at
lower redshift.

Figure 10 shows the median $R$ and confidence intervals in $R$ (corresponding
to $\pm 1 \sigma$) for all QSO pairs, as a function of $S$.
As discovered in Paper III, there is a significant trend of median
estimated $R$ with $S$, contrary to our assumed model.
The slope in a linear fit of $R$ versus $S$ is $0.43 \pm 0.08$ for all QSO
pairs.
If, noting that Q0107-0234/0107-0235 appears to be discrepant, one leaves it
out, one finds that the trend of $R$ with $S$ is almost unchanged and more
significant, with slope of $0.43 \pm 0.08$.
The other lower redshift QSO pair, Q1517+2357/1517+2356, falls {\it below} the
trend set by higher redshift QSOs (as does Q1026-0045A/B).

With the $R(S)$ dependence removed, one finds the Q0107-0234/0107-0235 point
sitting $2.3 \sigma$ above the minimum $\chi^2$ linear fit of $R$ versus $z$,
with Q1517+2357/1517+2356 $0.9 \sigma$ below, and Q1026-0045A/B $0.8 \sigma$
below.
There is no significant trend of size increase with $z$ (best
fit $\partial R/\partial z = 21$ kpc per unit $z$, with an error of 51~kpc per
unit $z$).
While Q0107-0234/0107-0235 suggests a trend of $R$ with $z$, this trend is not
supported by any other data, and Q0107-0234/0107-0235 alone is insufficient to
establish an effect.
Perhaps $R(z)$ changes more rapidly at smaller $z$ than at larger, but more
data from close, low redshift QSO pairs would be needed to substantiate this.

Paper III shows that there are at least three viable alternatives for
the dependence of $R$ on $S$: small scale clustering, elongated clouds
(filaments) and a non-uniform $R$ value among the clouds.
From QSO pairs alone, it is very difficult to distinguish which of these, or
which combination, is in play.
The triplet data, however, is used below to probe the shape of Ly $\alpha$
forest clouds.

\subsection{Elongated Absorbers}

It is also possible to use ``hit'' statistics to test directly the
non-spherical models.
For instance, we can ask if Ly $\alpha$ absorbers are elongated, by simulating
the projection of a simple shape against the sky in an isotropic collection
of orientations.
Paper III suggests elongation of the clouds into filaments as one of
several possible explanations for a dependence of inferred cloud size $R$ from
our Bayesian model on sightline separation $S$.
Furthermore, numerical models of intergalactic objects in the early
Universe (Zhang et al.~1995, Katz et al.~1995, Miralda-Escud\'e et al.~1996)
tend to find elongated structures on the scale of several hundred kpc as those
with properties most similar to Ly $\alpha$ clouds.
This should be contrasted to purely gravitational simulations e.g.~Shandarin
et al.~(1995), which tend to produce sheet-like structures first.
The triplet is ideal for determining whether the hit statistics deviate from an
$S$-independent $R$ due primarily to elongated clouds; single, long, thin
filaments are incapable of intercepting all three sightlines.
QSO triplets carry with them the potential to measure the aspect ratio of
filaments, $a = l/2R$, and $R$ independently (where $R$ is the cross-sectional
radius, and $l$ is the length of the filament).

One possibility that we do not discuss in this subsection are sheets or disks
($a < 1$), since their behavior in terms of hit statistics is similar to
spheres of slightly smaller radius (Paper III) in the case of sightline
pairs.
This is still true for triple sightline hits.
As $a\rightarrow0$, the utility of the triple sightline approach is to
distinguish the face-on projected shape of the disk, which we judge to be a
less interesting problem.
We discuss other test for disk-like structure in \S 4.7 and 4.8.
Here we pursue the following question: are the two- and three-way hit
statistics in the QSO triplet consistent with elongated,
circular-cross-sectioned rods of some radius $R$ and aspect ratio $a$?
This is motivated by the realization that long, thin filaments cannot span all
three sightlines (if the minimum distance across their triangle projected onto
the sky is larger than the width of the filament), whereas a circular cloud of
the same volume as the filament might easily do so.
Such an effect should be expressible as the probability of clouds of a given
shape and size hitting two or all three sightlines whenever they hit one.
This is accomplished by, first, measuring in the actual spectra's linelists the
probabilities $P_{ab}$, $P_{ac}$, $P_{bc}$ and $P_{abc}$, (defined as $P_{ab}$
being the probability of a line in A resulting in a hit in B, or vice versa,
and likewise for the other probabilities) and, secondly, simulating the same
probabilities by a numerical simulation of cylindrical rods of various $a$ and
$R$ values oriented in an isotropic distribution and intercepting (or not) two
or three sightlines with the same spacings as those between KP 76, 77 and 78
(sightlines ``A'', ``B'' and ``C'', respectively).

For this test, we need the largest sample possible, hence for $W_o > 0.4$\AA,
the $\langle z \rangle = 2.14$ sample.
As a preliminary indication, consider that of the 29 $W_o > 0.4$\AA\ hits, some
15.6 are expected at random.
Consider also the large number of ``multiple hits'' of two or three pairs
between all three sightlines involving the same Ly $\alpha$ clouds, at
$z = 1.938$ (three pairs involving
three lines in all three QSO spectra), $z = 2.042$ (three pairs, three lines,
three QSOs), $z = 2.113$ (three lines, two pairs, three QSOs), $z = 2.138$
(three lines, three pairs, three QSOs) and $z = 2.183$ (three lines, three
pairs, three QSOs), for a total of fourteen pairs.
In other words, the entire excess in $\Delta v < 200$~km~s$^{-1}$, $W_o >
0.4$\AA\ pairs might be due to these five groupings.
This, even by itself, argues for clouds that are not simply long, thin
filaments (in comparison to the sightline separations), since such clouds
cannot span the three sightlines.

The probabilities $P_{ab}$, $P_{ac}$ and $P_{bc}$ are computed by counting the
number of relevant pairs and dividing by the geometric mean of the number of
lines in each sightline's sample that is involved (23, 24 and 29 in KP 76, 77
and 78, respectively, for $W_o > 0.4$\AA), eliminating the fraction of pairs
that are expected at random (reduced to a fraction of 13.4/29 of the original).
Errors are computed from the Poisson distribution around a mean equaling the
actual number of observed hits.
For three-way hits ($P_{abc}$), since any given line has a 35\% chance of being
accidentally involved in a hit, the probability is reduced by this fraction.
Errors are computed to first order by considering the Poisson statistics for
the multiple hits (since they are likely not chance events) and in the sample
size, then adding in quadrature the error in false hits, leading to the 68\%
confidence interval (corresponding to $\pm1\sigma$) assigned to each
probability:
$P_{ab} = 0.30^{+0.19}_{-0.15}$,
$P_{ac} = 0.18^{+0.16}_{-0.12}$,
$P_{bc} = 0.06^{+0.14}_{-0.10}$, and
$P_{abc} = 0.16^{+0.14}_{-0.10}$.
Formally, $P_{bc}$ cannot be less than zero, nor smaller than $P_{abc}$; its
small value appears to be a result of an unusually small number of random hits
in BC.
If we compute the probabilities in a different way, by recognizing that triple
hits are almost certainly real (not accidental) and that the remaining number
of real hits must be non-negative, this produces a new set of probabilities for
the pairs of sightlines:
$P^\prime_{ab} = 0.24^{+0.20}_{-0.13}$,
$P^\prime_{ac} = 0.12^{+0.13}_{-0.12}$, and
$P^\prime_{bc} = 0.15^{+0.13}_{-0.11}$.
Normally (if the clouds are oriented isotropically and parcels of gas within
the cloud have a two-point correlation function that decreases monotonically
with separation), one should expect $P^\prime_{ac} \ge P^\prime_{ab} \ge
P^\prime_{bc}$.
The probability $P^\prime_{ac}$ disobeys this most significantly; we will
encounter this again below. 

As a comparison, we produce a model of a single, rotating, translating cylinder
(of circular cross-section) that is stepped in a fine grid (10 $h^{-1}$ kpc in
two orthogonal directions perpendicular to the sightline)
across three sightlines with the same spacing as the triplet.
The rod is ``hard-edged'' with no variation in $W_o$ over its projected shape.
The rod is made to point in 1280 isotropically distributed directions at each
grid point.
This simulation is done for rods with cross-sectional radii $R$ that are
positive multiples of 25 $h^{-1}$ kpc up to 500 $h^{-1}$ kpc (in proper
coordinates), and for aspect
ratios (length divided by diameter) of positive integral values up to 20.
For each rod shape, size, orientation and translation, the hit on one, two or
three sightlines, A, B, C, AB, AC, BC or ABC, is evaluated.
For each translational grid, the number of one-way hits are required to be all
equal, $N_a = N_b = N_c$.
The probabilities are computed for each rod shape and size by $P_{ab} = N_{ab}
/N_a$, ... , $P_{abc} = N_{abc}/N_a$.
Example contour plots (unsmoothed) of two of these probabilities, $P_{ab}$ and
$P_{abc}$, respectively, are shown in Figures 11a and 11b.
$P_{ac}$ and $P_{bc}$ resemble $P_{ab}$, qualitatively, while $P_{abc}$,
containing information about the shape, is more distinct, going to zero for
$2R < 406 ~ h^{-1}$ kpc, the minimum distance across the triangle described by
the triplet.
The contours at $P=0.01$ show a few ripples at the level of about 0.002,
indicating the degree of discreteness error in the model.

The contours for the calculated value for each probability $P_{ab}$,
etc.~(solid curves) and a confidence interval (68\% - dotted curves) is plotted
in Figure 12a.
The same for $P^\prime_{ab}$, etc.~is plotted in Figure 12b.
The probability $P_{abc}$ is inconsistent with the mean of the two-way
probabilities $P_{ab}, P_{ac}, P_{bc}$ at a level greater than $1\sigma$ for
any aspect ratio $a > 4$ (accounting for the errors in all $P$).
For $a\la 2$, in both plots, $P_{abc}$ is consistent with $P_{ab}$ and $P_{bc}$
(or $P^\prime_{ab}$ and $P^\prime_{bc}$) at a level less than $1\sigma$.
For $P_{ac}$ (or $P^\prime_{ac}$), however, the disagreement with $P_{abc}$ is
of the order of $1-2\sigma$ even for $a \la 2$.
As noted above, $P_{ac}$ is anomalously low relative to the other
probabilities, and this is reflected here.
Also, the statistical significance of the difference between large and small
$a$ is not great, decreasing from a maximum $1.2 \sigma$ for large $a$ to
0.6-0.7$\sigma$ for $a < 2$.
Nevertheless, all probabilities have best agreement for $1 < a < 3$ and
$198~h^{-1}$kpc$~ < R < 510~h^{-1}$kpc (larger $R$ at smaller $a$), where
$\sigma < 0.8$ (weighting the three two-way probabilities at two-thirds that of
$P_{abc}$).

The result that the longest dimension across these absorbers exceeds somewhat
$700~h^{-1}$kpc is guaranteed by the observation that there are significant
numbers of hits between the sightlines, plus the assumptions of the model.
The derived shape and size must span the sightlines; however, the way in which
it does so - by long, thin clouds or large near-spheres - depends on the
relative number of three-way hits.
We conclude that unclustered filaments alone are less likely to explain our
shape information on the $W_o > 0.4$\AA\ Ly $\alpha$ forest.
The absorbers are more likely to be nearly circular in cross-section (disks or
spheroids), or, if elongated, their hit behavior on the scale of hundreds of
kiloparsecs must be dominated by clustering of filaments, not the shape of the
filaments themselves.

Obviously, given Figure 6, no such $\Delta v \le 200$~km~s$^{-1}$ triple hits
exist for the $W_o > 0.8$\AA\ sample.
Expanding the velocity interval corresponding to a hit to $600$~km~s$^{-1}$,
the interval of possible $W_o > 0.8$\AA\ clustering seen in Figure 6, we find
one triple hit ($z = 2.05665$, $W_o = 0.86$\AA\ in KP76; $z = 2.05506$, $W_o =
1.24$\AA\ and $z = 2.05056$, $W_o = 1.77$\AA\ in KP77; $z = 2.06117$\AA\ in
KP78), being composed of four lines, not just three.
This feature is completely independent of the $\Delta v \le 200$~km~s$^{-1}$,
$W_o > 0.4$\AA\ triple hits above, and contributes one hit to the $P_{ab}$,
$W_o > 0.4$\AA\ signal.
It is spread over 1040~km~s$^{-1}$; this is about six times the Hubble flow
across the transverse dimension of the QSO triplet.
We should note, however, that we would expect approximately 2.1 such triple
hits at random in this sample, so there is little to be concluded from this
datum.

Likewise, even though there is a statistically insignificant excess in two-way
hits in the $0.2 < W_o \le 0.4$\AA\ sample; this might mask a more significant
three-way signal.
In truth, there is a slight deficit in such three-way hits compared to the
random expectation, so we can conclude little, except that $P_{abc}$ is likely
smaller for $0.2 < W_o \le 0.4$\AA\ than for the $W_o > 0.4$\AA\ sample.

\subsection{Consistency of Triplet Hits with Spherical Absorbers}

Are spheres of different $R$ required to explain $P_{ab}$, $P_{ac}$, $P_{bc}$
and $P_{abc}$?
Or is a distribution of spheres of different radii even consistent with the
data?
Figure 13 shows how the four probabilities vary for a given uniform $R$ value
common to the whole population of spherical clouds.
A single $R$ value should be consistent with the 95\% upper limits on all three
$P^\prime$ values and $P_{abc}$.
This implies an upper limit $R < 468~h^{-1}$~kpc from $P_{ac}^\prime$.
On the other hand, all of the two-way probabilities ($P_{ab},~P_{ac},~P_{bc},~
P_{ab}^\prime,~P_{ac}^\prime,~P_{bc}^\prime$) are roughly equal, a condition of
large ($R>600~h^{-1}$~kpc) clouds.
For such large clouds, however, $P$ values are much larger ($\ga0.3$).
In order for both the ratio in $P$ values and their rough magnitudes to be
satisfied by a distribution of spheres of varying radii, a sub-population of
$R>600~h^{-1}$~kpc clouds must be diluted by a larger portion ($\sim$50-70\% of
the total cross-section) of smaller clouds which do not span the sightlines and
hence do not contribute significantly to hit counts.
These reduce all $P$ values by the proportion between the cloud
sub-populations' total cross-section, but keep the ratios of various $P$ values
intact at their $R>600~h^{-1}$~kpc ratios.
Such a spectrum of cloud sizes is consistent with our previous constraints
(Paper III) on a power-law distribution of spherical cloud radii, which we do
not re-derive here.
We have not yet managed to challenge this hypothesis on the basis of measured
cloud parameters.

\subsection{Kinematics of Flattened Absorbers in the Hubble Flow}

While it is difficult with hit statistics to distinguish disks from spheres,
we can use the Hubble expansion to probe the probable shape of a cloud.
Starting as an object expanding nearly as fast as the rest of the Universe, the
absorber may collapse in one or two dimensions while still expanding in an
orthogonal one.
We can then distinguish a filamentary or sheet-like object by the tilt of the
direction of expansion relative to the line of sight, with one side expanding
toward the observer, while the other recedes.

The three sightlines of the triplet rest on a circle 189.7 arcsec in diameter,
separated in the sky by 90$^\circ$, 110$^\circ$, and 160$^\circ$ with
respect to the center, close enough to equilateral to always sense most of the
velocity shear across the circle for an expanding sheet.
For the five triple-hit objects, we find the best fit in magnitude and angle of
the line of nodes for this shear pattern, and find maximum velocities across
the circle's radius of 20, 35, 175, 65 and 95 km~s$^{-1}$ (to the nearest
5~km~s$^{-1}$) for the $z=1.938$, 2.043, 2.113, 2.138 and 2.183 objects.
respectively.

In comparison the Hubble expansion across this radius at $z \approx 2.1$ is
about 165~km~s$^{-1}$ (to within about 20\% for the cosmological models we
consider), whereas various inclination angles $i$ can project this to zero or
nearly infinite velocities.
Figure 14 shows the expected cumulative distribution of shear velocities for
a sheet expanding in the Hubble flow, for a random distribution of $i$ values,
and for our three cosmological models.
A mean measurement uncertainty of 20~km~s$^{-1}$ is folded into the $v_{max}$
distribution.

Figure 14 allows us to use a one-sample Kolmogorov-Smirnov (K-S) test to
determine whether the observed $v_{max}$ distribution is consistent with
theoretical expectation.
For the three cosmological models, (1, 0), (0.1, 0) and (0.1, 0.9),
respectively, the null hypothesis (consistency with Hubble expansion within
sheets or disks), cannot be rejected, at levels of 50\%, 70\% and 99\%, in the
sense that (0.1, 0.9) is more consistent.
The K-S test does not reject a model based on expanding sheets, while elongated
filaments encounter difficulty in \S 4.5.
Likewise, we cannot reject a simple gaussian distribution of velocities.
We need a few times as many such measurements to discriminate between these two
models.	

A further prediction of the sheet model is a correlation between $v_{max}$ and
$W_o$, which might be evident unless the perpendicular column density through
different absorbers or along different sightlines in the same
cloud shows scatter greater than about order unity.
Such a correlation should be a proportionality, or at least monotonic, hence
susceptible to a rank-order test.
We choose the median $W_o$ value of all lines contributing to the triple hit
objects.
The results from Spearman's Rank Correlation test shows that the $W_o >
0.4$\AA\ sample is consistent with a monotonic $v_{max}(W_o)$ at the 72\%
confidence level.
When the $W_o > 0.8$\AA\ triple hit is included, however, the confident level
becomes 93\%, suggestive of the tilt of the absorber being an important
parameter.
We should note that when the same test is applied to two-way hits, however, no
such correlation is seen.
This may suggest that three-way hits must be required to bring this correlation
out of the noise, or that the objects which span the three sightlines are
sheet-like.

\subsection{Other Tests of Ly $\alpha$ Absorber Shape, Size and Clustering}

In the Introduction we review the several recent theoretical works proposing
observational tests involving pairs of sightlines.
Most of these can be applied to the current data, and we consider them in turn.

Many of these tests involve figures published in the four theoretical works.
While we present our data here in a form which can be compared most directly to
these other results, we avoid reproducing all of their relevant figures, and
refer the reader to the original papers (Cen \& Simcoe 1997,
Charlton et al.~1995, 1997, Miralda-Escud\'e et al.~1996).
Additionally, Charlton et al.~(1997) discuss other kinematical tests similar to
those considered in the prevous sections, which we will not rediscuss here.

\subsubsection{Correlated Flux between Sightlines (Miralda-Escud\'e et
al.~1996)}

Miralda-Escud\'e et al.~(1996) consider the correlation of Ly $\alpha$
absorption as a function of transverse separation between adjacent sightlines.
This is expressed purely in terms of the correlation of the transmitted flux
between sightlines, not correlated line detections as in this paper.
They define a correlation coefficient $\xi_f (\Delta v, \Delta r)$ ({\it not} a
two-point correlation function as usually defined) which describes the
correlation between the transmitted flux $F$ in two adjacent sightlines:
$\xi_f (\Delta v, \Delta r) =
\langle [ F (r, v_0) - \langle F (r, v) \rangle_v ]
[F(r+\Delta r,v_0+\Delta v) - \langle F (r+\Delta r, v) \rangle_v ] \rangle_r /
\{ \langle F^2 \rangle_{v,r} - \langle F \rangle_{v,r}^2 \}$,
where for instance, $\langle F (r, v) \rangle_v$ refers to the expectation
value of $F$ along a sightline at location $r$ on the sky, averaging over the
sightline (which is parameterized by velocity $v$ along the sightline.
Note that $\Delta r$ corresponds to our $S$.)~
Necessarily, $-1 < \xi_f < 1$ and $\xi_f \rightarrow 1$ for
$\Delta v\rightarrow 0$ and $\Delta r\rightarrow 0$.
They show (in their Figure 13) that $\xi_f (\Delta v, \Delta r)$ drops to about 0.5 of its peak ($\Delta v=0$) value at $\Delta v \approx 60-120$~km~s$^{-1}$
for various $\Delta r$,
and drops close to zero for $\Delta v \ga 250$~km~s$^{-1}$ (less than 0.1 for
$\Delta v \ga 130$~km~s$^{-1}$), regardless of $\Delta r$.
The peak at $\Delta v =0$ falls to 0.5 at $100~h^{-1}$kpc (proper
separation) and to 0.17 at the largest $\Delta r$ shown, $418~h^{-1}$kpc.
While this prediction does not quite extend to $\Delta r$ values for the KP
triplet, 496 to 720 $h^{-1}$kpc, it is reasonably securely extrapolated to
$\xi_f \approx 0.06$ at $\Delta r = 599 h^{-1}$kpc, the mean for the triplet.
All of these values apply at $z=3$.

Since we do not resolve most of the Ly $\alpha$ forest, we cannot compute
$\xi_f$ directly.
We take this opportunity to note that analysis of theoretical models of the
high redshift neutral hydrogen distribution should continue to consider the
alternative approach of correlated line detections.
Here and in many potential cases in the future, QSOs in close pairs are
sufficiently faint that lines can be detected but not usefully resolved, even
with 8-10 meter class telescopes.
Nevertheless, we can compute the statistical moments of $F$ e.g.~$\langle F
(r, v) \rangle_v$ and $\langle F^2 (r, v) \rangle_v$ since we also have high
resolution Keck HIRES data for KP 77 (Crotts et al.~Tytler 1997), finding
$\langle F \rangle_v = 0.725$ and $\langle F^2 \rangle_v = 0.807$.
(Note that we cannot take the expectation over $r$, of course.)~
For the $W_0 \ge 0.4$\AA\ lines in this sample, not all covered by the HIRES
data, we attempt to translate our data on absorption lines into measures of the
correlation of actual flux, treating Ly $\alpha$ forest spectra as continuous
functions in wavelength rather than discrete lines.
This is accomplished by considering only the lines with $W_0 \ge 0.4$\AA, since
the rest are uncorrelated.
These lines are replaced with Voigt profiles of the same $W_0$ and an assumed
value of the Doppler parameter $b = 30$~km~s$^{-1}$.
(The result is fairly insensitive to the adopted $b$ value.)~
This results in the value measured from our $\langle z \rangle = 2.14$, $W_0
\ge 0.4$\AA\ KP sample of $\xi_f (\Delta v=0, \Delta r=599~h^{-1}$kpc$) =
0.069$, with $1\sigma$ errors of about 0.01.
This value is consistent with the theoretical $\xi_f \approx 0.06$.

When we attempt the same calculation for the close pair 1343+2640A/B at
$\langle z \rangle = 1.86$, we find
$\xi_f (\Delta v=0, \Delta r=40~h^{-1}$kpc$) \approx 0.40$, with $1\sigma$
errors of about 0.05.
This should be compared to a model prediction of about 0.8.
This difference is due in part to the fact that we have ignored weaker lines,
but even in 1343+2640A/B these are more weakly correlated, as we consider in
the Discussion section.
Most likely the measured value remains smaller than predicted, most likely
$\xi_f (\Delta v=0, \Delta r=40~h^{-1}$kpc$) \la 0.6$.

\subsubsection {Fraction of Coincident Lines versus $\Delta r$ and $\Delta v$
(Cen \& Simcoe 1997)}

Cen and Simcoe (1997) use the same simulation as Miralda-Escud\'e et al.~(1996)
to study the nature of individual clouds of H I, over redshifts $2 < z < 4$.
They study clouds as defined by regions isolated by different threshold
baryonic density cuts expressed in terms of mean baryonic density $\rho(x) /
\langle \rho \rangle > \rho_{cut} = 3$, 10 or 30, values chosen by the authors.
They find clouds that are relatively round and small (mean radii $\approx$
23~$h^{-1}$~kpc for $\rho_{cut} = 10$ and 33~$h^{-1}$~kpc for $\rho_{cut} =30$,
and commonly with axis ratios of about 1:2:4 or, a smaller fraction of the
time, closer to spherical).
They also argue that on scales larger than these mean diameters any observed
hits in adjacent sightlines are due to clustering of clouds, not cloud
structure itself.
Nevertheless, it is clear from the contours at lower $\rho$ that larger, more
sheet-like or filamentary chains of clouds are also present in the simulation,
on scales up to nearly the simulation box size of 2.5~$h^{-1}$~Mpc (proper)
at $z=3$.
For comparison with observations, they state that $\rho_{cut} = 10$ in their
simulation corresponds to $N_{HI} = 1.1\times10^{14}$cm$^{-2}$, or $W_o =
0.29$\AA\ for $b=30$~km~s$^{-1}$, which together with the corresponding value
for $\rho_{cut} = 30$, $W_o = 0.47$\AA, straddles our limit $W_o = 0.4$\AA.

In close analogy to our Figure 10 and Table 8, Cen \& Simcoe present their
Figure~9, which describes the fraction (compared to all lines) of lines
coincident between QSO sightlines as a function of proper transverse separation
$\Delta r$ and velocity ``hit'' window width $\Delta v$.
We find that the line correspondence ratio (corresponding to their Figure 9's
ordinate) is
$N_{co}/N_{tot} = 0.93 \pm 0.06, 0.29 \pm 0.17, 0.40 \pm 0.11, 0.57 \pm 0.13,
0.18 \pm 0.06, 0.38 \pm 0.07, 0.27 \pm 0.07$ and $0.22 \pm 0.06$, respectively,
for the pairs listed in Table 8, in order of increasing separation, with
r.m.s.~binomial error shown.
This is computed considering that a hit corresponds to two lines in the sample.
A comparison with the $\Delta v = 150$~km~s$^{-1}$, $\rho_{cut} = 10$ or 30
curves from Cen \& Simcoe Figure~9 shows that all pairs, with the possible
exception of Q1026-0045A/B and the LB pair (with $N_{co}/N_{tot} = 0.18$ at
$\Delta r = 432~h^{-1}$~kpc, which falls off of the graph in $\Delta r$), lie
at least one sigma above the highest corresponding theoretical curve.
(The KP points are also off of the graph, but seem to lie at least one sigma
above, as well.)~
While we use $\Delta v = 200$~km~s$^{-1}$ in Table 8, the effect of this over
$\Delta v = 150$~km~s$^{-1}$ is small compared to the difference between
theoretical and measured results.
Furthermore, Cen \& Simcoe's Figure~9 applies to $z=3$, but their Figure~10
shows that $N_{co}/N_{tot}$ does not grow at all between $z=3$ at the typical
redshifts $z \approx 2$ of the sample in Table 8.
The simulated absorption lines are less correlated between sightlines than the
observed ones.

This general result for $N_{co}/N_{tot}$ is consistent with the small $\xi_f
(\Delta v=0,\Delta r=40~h^{-1}$kpc$)$ result of Miralda-Escud\'e et al.~(1996),
which is not surprising given their use of the same model.
Casting this in terms of the small clouds delineated by Cen \& Simcoe, one 
tends to conclude that more power is needed in their model on wavelengths of
$\sim 100-1000~h^{-1}$~kpc, which is slightly smaller than the proper size of
the simulation volume of 2.5~$h^{-1}$~Mpc at $z=3$.

\subsubsection
{Spatial Clustering of Ly $\alpha$ Absorbers (Cen \& Simcoe 1997)}

Cen \& Simcoe plot the cloud two-point correlation function $\xi (r)$, where
$r$ is the comoving separation between absorbers, in their Figure 14.
Our measurement of clustering in the KP triplet at separations $\langle S
\rangle = 599~h^{-1}$~kpc, plus a line-of-sight component $\Delta v =
200$~km~s$^{-1}$ yields a typical proper separation of $625~h^{-1}$~kpc.
At $z=3$ this produces $\xi = 0.19$ and 0.27, for $\rho_{cut}=10$ and 30,
respectively.
Our measurement of $\xi$ was made at a different redshift $\langle z \rangle =
2.14$, but Cen \& Simcoe's Figure 15 allows us to account for the evolution of
clustering power from $z=3$ to 2.14, between which correlation length $r_0$
increases by about 10\%.
Given a correlation function $\xi \propto r^{-1.8}$, the value of $\xi$ at a
proper separation of $625~h^{-1}$~kpc should increase by about 18\%.
Since we measure a value at this separation of $\xi = 1.88^{+0.78}_{-0.50}$,
we find that the $\rho_{cut}=30$ model result and actual measurement are
inconsistent at about the $3 \sigma$ level.
This may also be due to the lack of longer wavelength modes in the clustering
power spectrum, as Cen \& Simcoe also speculated.

\subsubsection {Fraction of Coincident Lines versus $\Delta r$ and $\Delta v$
(Charlton et al.~1997)}

Like Cen \& Simcoe, Charlton et al.~(1997) consider model simulations (Zhang et
al.~1995) to construct statistical measures which can be compared to
observations.
Also like Cen \& Simcoe, they discuss how the fraction of common lines varies
with sightline separation $S$ (their $D$).
Their Figure 2e corresponds most closely to our sample, with $z=2$ and a curve
at $N_{HI} = 10^{14}$cm$^{-2}$.
This value of $N_{HI}$ is slightly smaller than our $W_0=0.4$\AA\ cutoff for
typical $b$ values.
The values of $N_{co}/N_{tot}$ (their $f_{co}$) listed above for $z \approx 2$
samples (excluding 0.57 for 0107-0234/35) all scatter within $1\sigma$ of the
theoretical curve, except for 1517+2356/57, which is too low, and the adjacent
value, 1623+2651A/B, which is too high.
All six QSO pairs, taken together, are consistent with this curve and have
residuals that are might arise from a reasonable $\chi^2$ distribution.

\subsubsection
{Median Absorber Size Implied from Hit Statistics (Charlton et al.~1997)}

Charlton et al.~(1997) study how the size of clouds implied by hit statistics 
change with pair separation, for the structures within the numerical model
(Zhang et al.~1995).
This implied size depends on whether one assumes as a working model for the
clouds a shape of spheres or thin disks.
Figure 3 of Charlton et al.~(1997) corresponds closely to our Figure 10.
Unfortunately, none of their sub-figures correspond exactly, but their
Figure 3f is a close match, using $\Delta v = 150$~km~s$^{-1}$ instead of
200~km~s$^{-1}$, and disks instead of spheres.
From Paper III, however, results for spheres can be converted to disks by
multiplying by a factor of about 1.5.
With this adjustment, all $z \approx 2$ pairs fall within about $1\sigma$ of
the theoretical curve, with the small $S$ curve slightly undercutting the
observed values.

\subsubsection {Linestrength Correlation between Sightlines (Charlton et
al.~1995, 1997)}

In Charlton et al.~(1995) a co-distribution of column densities ($N_a$ and
$N_b$) for adjacent sightlines $a$ and $b$ of various separations is considered
for various idealized disk-shape cloud models, whereas in Charlton et
al.~(1997) the same test is applied to the model simulation of Zhang et
al.~(1995).
In the first case, they consider separations up to the cloud diameter, with the
best discrimination occurring for separation less than about the cloud radius.
In the second paper, they consider proper separations up to 200~$h^{-1}$~kpc.
Given the separation range of the theoretical effect, 1343+2640A/B and
0307-1931/32 are most valuable comparison among observed pairs.
We will also consider the KP triplet.
Following Paper III, it is wise to consider only those lines thought to be
unlikely to be contaminated by interloping metal lines, and with a $S/N=3.5$
cutoff imposed.

Again, our data are expressed in terms of $W_0$, not $N_{HI}$, so we must
assume a value for $b$ (of 30~km~s$^{-1}$), which will introduce scatter into
the transformation between $W_0$ and $N_{HI}$.
Fortunately, Charlton et al.~(1997) also recompute the effect in terms of the
difference of equivalent widths $|W_a - W_b|$ versus the strength of the
strongest line $max(W_a,W_b)$ (their Figure 5) which is directly comparable to
Figure 2b of Paper III (except for the two distributions corresponding to
slightly different $S$ values, 40~$h^{-1}$~kpc for the observed pair
1343+2640A/B and 50~$h^{-1}$~kpc for model).
The tight theoretical correlation is supported by the observed values, with
none of the observed $W_0$ locii for uncontaminated lines being inconsistent
with the theoretical result, and with the potentially contaminated lines also
being in reasonable agreement.
As Charlton et al.~(1997) note, a larger data set would be desirable.

We also consider the pair 0307-1931/32 at $S = 231~h^{-1}$~kpc separation.
Figure 15a shows the distribution of $|W_a - W_b|$ versus $max(W_a,W_b)$ for
0307-1931/32, analogous to Figure 2b of Paper III for 1343+2640A/B.
(Since we could not reanalyze this spectrum, we do not attempt to remove Ly
$\alpha$ potentially contaminated by superimposed metal lines.)~
In general, the points in Figure 15a must lie below the $|W_a - W_b| =
max(W_a,W_b)$ diagonal, and the farther they fall from this line, the more
homogeneous they are.
Unlike the idealized models of Charlton et al.~(1995), with analogous results
presented in their Figure 1, there is still a close agreement between the
observed wide-pair locus and the locus predicted for very close pairs
e.g.~Charlton et al.~(1997) Figure 5.
We draw no interesting conclusion from this comparison.

The results for $|W_a - W_b|$ versus $max(W_a,W_b)$ are more interesting for
the KP triplet, as shown in Figure 15b.
Due to the density of points, we do not present error bars, which are typically
much less than 0.1\AA\ (hence about the size of the point symbols).
Each point represents a Ly $\alpha$ forest line coincidence ($\Delta v < 200$
km s$^{-1}$) free of probable
metal-line contamination (from systems at different redshifts) and without
associated metal lines (from the same redshift system).
This figure's distribution of points seems to fall at least as far below the
$|W_a - W_b| = max(W_a,W_b)$ line as those in Charlton et al.~(1997) Figure 5
for the close pair 1343+2640A/B.
The points flagged by the three-legged crosses denote absorbers spanning all
three KP sightlines (with $W_0 > $0.4\AA).
While this minimum $W_0$ selection guarantees that the points sit at least
0.4\AA\ below the $|W_a - W_b| = max(W_a,W_b)$ line, it is significant that
those objects which span all three sightlines appear to be the most uniform of
any in the $S/N \ge 3.5$, uncontaminated, Ly-$\alpha$-only sample.
Separate from this selection effect, one can still state that the most
homogeneous, strong lines also span all three sightlines.
This argues that these compose a well-defined class of objects and are not
produced by random superposition.
Furthermore, the uniformity in $W_0$ argues against these pairs being produced
simply by clustering of smaller objects, but indicates some coherence in the
H~I distribution on 0.5-0.8~$h^{-1}$~kpc scales, such as sheets of gas, and not
filaments on these scales.

\subsubsection {Distribution of $N_{HI}$ among Anti-coincident Lines (Charlton
et al.~1995, 1997)}

Charlton et al.~(1995) present the shape of the distribution $f$ of column
densities of lines {\it not} participating in hits within close pairs
($N_{ac}$) as a sensitive discriminator of cloud shape e.g.~disks versus
spheres.
This difference appears most strongly at the high $N_{HI}$ end, where spheres
show a sharp cutoff at a position dependent on the size of the clouds relative
to $S$, whereas disks show a gradual reduction at all $N_{HI}$ values, with
only a slow change in the slope of $log(f)$ versus $log(N_{ac})$.
Charlton et al.~suggest this test for a set $N_{HI}$ cutoff in the linestrength
of the ``missing'' line.
(Lines weaker than this can produce anticoincidences in neighboring
sightlines.)~
We impose a cutoff $W_o > 0.3$\AA, which corresponds to $log (N_{HI}/$cm$^{-2})
= 14.16$ for $b=30$~km~s$^{-1}$.
One then studies the distribution $f$ of $N_{HI}$ in the remaining line.
We plot this function in Figure 6 for 0307-1931/32, 1517+2356/57 and the KP
triplet.
The distributions are renormalized to have the same number of lines at
$log (N_{HI}/$cm$^{-2})=13.5$, below which the sample is grossly incomplete.
(1343+2640A/B has so few anti-coincident lines that $f$ is poorly defined.)~
This results in a total of 18 anticoincident lines in 0307-1931/32, 17 in
1517+2356/57, and 41 from sightline pairs among the KP triplet.

The behavior of the anticoincident $N_{HI}$ distribution in Figure 16 differs
from that in Figure 2 of Charlton et al.
The observed distribution $f$ is broader and present at stronger $N_{HI}$ for
closer QSO pairs than in wider pairs, opposite the behavior predicted for
spheres or disks.
It is less discrepant with the more gradual increase in strength with
increasing $S$ found in disks.
It is also less discrepant with the behavior of lines in simulations (Charlton
et al.~1997, Figure 7), but is not in good agreement.

This puzzling behavior may simply be due to small number statistics.
The critical difference in the shape of $f$ for disks versus spheres occur over
the high $N_{HI}$ tail containing only 1-10\% of the lines.
We need several times as many data to adequately test these predictions.

\subsection{Metal-line Absorber Clustering}

The data on these five QSOs, especially the KP triplet, now includes enough
space probed with closely-spaced sightlines through the absorber distribution
that one can begin to ask how C~IV absorbers cluster in space, as measured
by the cross-correlation of the absorber distribution between sightlines, as
opposed to auto-correlating the distribution along single sightlines.
The latter approach, which has been presented in many works, carries with it
the danger that spatial clustering signals may be mixed with velocity
correlations due to internal velocity splittings within isolated absorbing
objects.
This is circumvented in the case of sightline cross-correlations; multiple
absorber redshifts per object simply are reflected in an increase in the number
of correlation pairs equally on all scales, not just for small separations
such as those internal to an absorber.
Such an increase in pairs cancels in the two-point correlation function $\xi$.

Figure 17 shows the number of cross-correlation pairs, binned in relative
velocity (as in Figure 6 for Ly $\alpha$ clouds) but in this case for all C~IV
absorbers identified for the KP triplet.
The solid curve shows all such C~IV pairs, and the dotted curve only those
pairs involving both absorbers in the ``probable'' or ``definite'' categories,
with detected C~IV $\lambda$1548.
The dashed curve in Figure 17 shows the same information, but for all C~IV
absorbers in the LB pair, Q0307-195 pair and KP triplet (where each sightline
is cross-correlated only with its partners in the same pair or triplet).
Q1343+2640 is excluded since its separation may be smaller than the size of
individual C~IV galaxy cross-sections, and Q0107-025 is at much lower redshift.
The first 500~km~s$^{-1}$ bin in Figure 16 shows more pairs than any other
except one (which shows nine at 14000~km~s$^{-1} < \Delta v <
14500$~km~s$^{-1}$).
Furthermore, the first 2000~km~s$^{-1}$ interval shows more pairs than any
other 2000~km~s$^{-1}$ interval in the entire $\Delta v$ range.
The average number of pairs per 500~km~s$^{-1}$ bin is 1.97, so the Poisson
probability of the first bin having the observed six pairs (or more) is 1.5\%,
and the probability of the first four bins having the observed 15 pairs is
also 1.5\%.
The probability of the second observation given the first is 14\%, so the
signal in the first bin is most significant, as long as we consider the first
bin to be uniquely special, a priori.
(If we do not, there are approximately 100 such equivalent bins, so the
probability of such a random occurrence in some bin is nearly unity.)
There is no obvious substructure in $\Delta v$ within the first bin, but the
contribution from the triplet is due to a single cluster at $z \approx 2.243$.
Specifically, the signal arises in the LB pair at $\Delta v = 233$~km~s$^{-1}$
and $z = 0.7376$, in the Q0307-195 pair at $\Delta v = 306$~km~s$^{-1}$ 
and $z = 2.0337$, and in the KP triplet at $\Delta v = 100, 261, 360,
490$~km~s$^{-1}$ at redshifts 2.2419, 2.2431, 2.2437 and 2.2451, respectively.
This is the most statistically significant signal ever seen showing that C~IV
clustering exists on small scales within sightline cross-correlations.
It corresponds to a two-point correlation function $\xi = 2.05^{+1.82}_{-1.21}$
(68\% confidence limits, as for the Ly $\alpha$ $\xi$ in \S 4.3), over proper
separations of 220 to 720~$h^{-1}$~kpc, approximately (for $q_o = 1/2$),
neglecting the distance component along the line of sight.
Including this component, it corresponds to distances up to about
1.05~$h^{-1}$~Mpc.
With $q_o=1/2$, this would correspond to velocity differences as large as
690~km~s$^{-1}$ in the Hubble flow.

While the statistical significance of this C~IV cross-correlation result is not
large, it should be compared to the single-sightline auto-correlation function
(Sargent, Steidel \& Boksenberg 1988), which shows, for single-sightline
splittings of 200~km~s$^{-1} < \Delta v < 600$~km~s$^{-1}$ a two-point function
of variously $\xi = 5.7 \pm 0.6$ or $\xi = 11.5 \pm 1.3$ depending on whether
or not a cutoff of $W_o = 0.15$\AA\ is imposed on C~IV$\lambda$1548 (in the
second case), with both cases excluding lines within 5000~km~s$^{-1}$ of the
QSO emission redshift.
The first case is inconsistent with our data at the $2\sigma$ level, the second
being even more inconsistent.
Our sample is more heterogeneous, not being defined by a specific cutoff on
$W_o$ or $\Delta v$ with respect to the emission redshift, but of the 38
C~IV$\lambda$1548, only 8 have $W_o < 0.10$\AA\ and 14 have $W_o < 0.15$\AA.
None of the lines contributing to the $\xi$ signal are within 5000~km~s$^{-1}$
of the emission redshift.
Our sample is closer to the second sample from Sargent et al.~(1988), and
is inconsistent with the results obtained from it.
This suggests that some of the power in $\xi$ seen by Sargent et al.~is not
due to spatial clustering, but to internal motion within absorbers.

Measuring structure on larger scales by cross-correlation of sightlines is
interesting, and unprecedented for unbiased samples (except for the full-sky
cross-correlational study of Crotts [1985] and the recent paper Williger et
al.~1996).
Tytler et al.~(1987), Heisler, Hogan \& White (1989) and Quashnock, Vanden
Berk \& York (1996) suggest structure up to 100~$h^{-1}$~Mpc comoving scales in
their studies of single-sightline metal-line autocorrelations.
This is consistent with the power we see at $\Delta v
\approx14000$~km~s$^{-1}$, but we would need a larger sample to be sure.
In general, however, the cross-correlational technique is most interesting on
smaller scales, where internal velocities may be important.
(Or it may be interesting for obtaining very dense, random samples, as
suggested for the Ly $\alpha$ forest in \S 4.2, or for studying certain
pre-selected anomalies e.g.~in the region around a close damper Ly $\alpha$
absorber pair: Francis et al.~1996, and the $\approx 100$~Mpc absorber cluster
in the Tololo QSO sample near 1037-27: Ulrich \& Perryman 1986, Jakobsen et
al.~1986, Jakobsen, Perryman \& Cristiani 1988, Robertson 1987, Cristiani,
Danziger \& Shaver 1987, Sargent \& Steidel 1987, Jakobsen \& Perryman 1992,
Dinshaw \& Impey 1996.)~
We are in the process of acquiring more such QSO pair data to study the
clustering properties of C~IV absorbers on Mpc scales.

\section{Discussion}

Returning to the clustering behavior of Ly $\alpha$ forest, one should note
that in both their contributions to large-scale and small-scale structure, the
$W_o > 0.4$\AA\ Ly $\alpha$ sample and weaker lines differ in their behavior.
Both in the small-scale two point correlation function, and in the presence of
large-scale underdensities, the $W_o > 0.4$\AA\ sample reveals more significant
features (see Figures 6 and 7) which are washed out in larger, lower $W_o$
cutoff samples.
This might lead one to believe that these two populations are distinct.
There is some evidence contradicting this, however, in that our original data
from Q1343+2640A/B includes some lines weaker than $W_o = 0.4$\AA, and these
also show signs of being associated with large absorbers.
Whereas the entire $S/N > 3.5 \sigma$ sample shows 11 hits and four misses,
the $W_o > 0.4$\AA\ subsample (see Table 8), once removed, leaves $N_h = 3$ and
$N_m = 4$ for the remaining $W_o < 0.4$\AA\ lines (typically 0.2\AA$< W_o 
<$0.4\AA, albeit not complete to $W_o = 0.2$\AA).
This implies that these weaker absorbers have a median predicted radius $R=
63~h^{-1}$~kpc (assuming unclustered, constant radiused spheres, and with 95\%
confidence bounds $35 ~h^{-1}$~kpc~$< R < 146~h^{-1}$~kpc), which, while
possibly smaller than the size in the $W_o > 0.4$\AA\ sample, still indicates
absorbers that are much larger than the visible size of galaxies, and within
about a factor of 2.5 of the size of low column-density H I galaxy halos at low
$z$ (Lanzetta et al.~1995).
Compared to C~IV absorbers as well (Steidel 1991, Paper I), these are still
large absorbers.

It is apparent, however, that the $W_o > 0.4$\AA\ absorbers are large enough,
at least in two dimensions, to span the gap between the triplet sightlines,
while the weaker lines may not.
It is interesting then, at this point, to examine the differences between those
objects spanning all three sightlines and those $W_o > 0.4$\AA\ clouds which do
not.
Of the five $W_o > 0.4$\AA\ triple-hit objects, we note that two are associated
with C~IV absorbers, while another probably is.
Furthermore, of the two that are not, one occurs at the same redshift of KP 79,
the $z=2.183$ QSO some 1.5~$h^{-1}$~Mpc away from the triplet sightlines (in
proper coordinates).
In the rest of the $W_o > 0.4$\AA\ Ly $\alpha$ sample, only 17 out of 144
absorbers have detected C~IV absorption.
These two populations are discrepant at the $3\sigma$ level.
Absorbers with strong C~IV may be associated with larger objects than strong
Ly $\alpha$ absorbers without strong C~IV.
The average velocity spacing between these groups of lines corresponds to
35~$h^{-1}$~Mpc (with $q_o=1/2$), roughly the scale of sheets between the voids
in the $z\approx 0$ galaxy distribution.
This correspondence echoes a suggestion long ago by Oort (1981) that Ly
$\alpha$ absorbers correspond to superclusters.
These data suggest something rather different, that a small subsample may
detect sheet-like structures on such scales.
It is not clear to us that such structures are evident the current
numerical simulations of the Ly $\alpha$ forest, nor that they necessarily
should be given the small volumes simulated by these models.

\section{Conclusions}

The strongest results indicated here are that the redshift correspondences
between Ly $\alpha$ absorbers in closer sightline pairs persists for proper
separations up to 0.5-0.8~$h^{-1}$~Mpc for lines stronger than $W_o = 0.4$\AA.
There is an indication that the shape of the clouds responsible for this signal
is not very elongated (aspect ratio $a < 4$, probably).
There is weaker evidence, however, of expansion of these clouds with the Hubble
flow in a way consistent with sheets or disks, and evidence of sheets of
relatively uniform gas density spanning 0.5-0.7~$h^{-1}$~Mpc transverse
separations.
This suggests that at least a large fraction of high column density absorbers
arise in expanding sheets.
While we do not have such a measurement for weaker lines, there is data from
Q1343+2640A/B (Paper I) that indicates that 0.2\AA$ \la W_o < 0.4$\AA\ clouds
are probably at least 35~$h^{-1}$~kpc in radius (95\%) confidence and probably
closer to 63~$h^{-1}$~kpc in radius, still very large compared to the luminous
size of galaxies and probably C~IV absorbers.

Despite the strong leverage in redshift with the inclusion of the LB pair and
Q0107-0234/0107-0235 and Q1026-0045A/B, the evidence for any evolution on the
size of $W_o > 0.4$\AA\ Ly $\alpha$ absorbers is not significant.

Our theoretical models to which the small-scale structure data are compared
here are intentionally crude.
Detailed comparison with numerical models incorporating
hydrodynamics and ionization, as well as gravity, at high redshift lead to
poor agreement in some cases and better in others, perhaps due to absence of
clustering power on scales comparable to the simulation box size.
It would appear that our current data may be inconsistent with the long
filaments ($a \ge 10$) that are produced in these models, but detailed
comparison of model and observation on as close a corresponding, quantitative
basis as possible are required to minimize the systematic errors involved in
analyzing the two forms of data differently.

C~IV absorbers are shown to cluster in space as well as velocity on small
scales ($\la 1~h^{-1}$~Mpc) for the first time in this paper, and while this
signal is weak, it seems inconsistent with the same sort of measurement via
single-sightline two-point correlation measurements.
The simplest explanation for this difference is additional splittings within
the absorbers themselves on velocity scales up to 600~km~s$^{-1}$.
This is hard to explain as the internal motions within galaxy halos (Sargent
et al.~1988), but may indicate large velocity flows due to non-gravitational
acceleration of gas, perhaps by shocks caused by star formation processes, 

With the exception of strong clustering seen on sub-Mpc scales, and some
indication of large, smooth sheets of gas,
none of the new conclusions from these data is stronger than about $3\sigma$,
and many more indications are less certain than this.
This calls for more QSO sightline pairs on the scale of separations less that
about 1~$h^{-1}$~Mpc proper separation.
Since most of these will be at lower redshift than even the KP triplet, it
also calls for UV sensitive instrumentation on large telescopes.
(However, if the behavior of only stronger lines is non-random, higher $S/N$
data will reveal no new effects.)
In the long term, both practical developments seem likely, but are the limiting
factors at present.

\acknowledgments

We thank Richard Green for providing the Ly $\alpha$ forest data on LB 9605 and
9612 published in Elowitz et al.~1995.
We very much appreciate the advice of Donald York in interpreting the complex
absorption line spectra of these objects.
We acknowledge the forethought of Foltz et al., who arranged for the {\it HST}
observations of LB 9605, 9612.
We would like to acknowledge support of the NSF under grant AST 90-22586, and
express our gratitude to the David and Lucile Packard Foundation for A.C.'s
fellowship.

\clearpage

\newpage 

\noindent Figure 1: the spectrum (in units proportional to photoelectrons as a
function of wavelength in Angstroms) for Q1623+2651A (KP 76).
The detected absorption lines are indicated by tickmarks extending downwards
from the counts$=$0 line, and the dashed line indicates the standard deviation
in counts per wavelength bin (typically 0.76\AA/bin in the top two panels and
0.72 in the bottom panel).

\noindent Figure 2: the spectrum for Q1623+2653 (KP 77), shown as for Figure 1.

\noindent Figure 3: the spectrum for Q1623+265B (KP 78), shown as for Figure 1.
 
\noindent Figure 4: the spectrum for 1517+2357 (LB9605), flux calibrated in
units of erg~s$^{-1}$~cm$^{-1}$, but otherwise shown as for Figure 1.
The wavelength interval per bin ranges from 0.36\AA\ below 2230\AA, to 0.51\AA\
for 2230\AA$<\lambda<$3270\AA, to 0.49\AA\ for 3270\AA$<\lambda<$3594\AA, to
0.76\AA\ above 3594\AA.
 
\noindent Figure 5: the spectrum for 1517+2357 (LB9612), shown as for Figure 4.
 
\noindent Figure 6a: the Ly $\alpha$ forest two-point velocity
cross-correlation function for all three sightline pairs among KP 76, 77 and
78, plotted as a histogram of the number of pairs in 50~km~s$^{-1}$ bins versus
velocity difference $\Delta v$.
The solid bars indicate pairs where both Ly $\alpha$ lines are stronger than
rest equivalent width $W_o = 0.8$\AA.
The densely shaded, lightly-shaded and unshaded bars show the same function for
samples with $W_o > 0.4$\AA, 0.2\AA\ and 0.1\AA, respectively.
The random pair count level for each of the $W_o$ levels, as $\Delta v
\rightarrow 0$, is indicated by the four dark tickmarks on the left edge of the
graph.
 
\noindent Figure 6b: as in Figure 6a, but for LB9605 and LB9612.
 
\noindent Figure 7a: the number of Ly $\alpha$ forest lines stronger than $W_o
= 0.4$\AA found in 15, 30 and 45~$h^{-1}$~Mpc wide bins (bottom solid curve,
dashed curve and top solid curve, respectively) along the combined sightlines
to KP 76. 77 and 78.
The bins are stepped in redshift by 1/4 of each bin width.
Note the depressions in line counts at $z \approx 2.08$ and $z \approx 2.37$,
referred to in the text and indicated by the dark tickmarks.
 
\noindent Figure 7b: the same as in Figure 7b, but for lines stronger than
$W_o = 0.1$\AA.
 
\noindent Figure 8: the foreground proximity effect for each QSO in the KP
triplet, expressed as the number of lines per unit $z$ as a function of
redshift.
The crosses shown the $z$ bin width and the $1\sigma$ errors in the number of
lines per bin as a function of mid-bin $z$.
The dashed slanted line shows the mean number of lines per unit z, $n_\gamma
(z)$, from a large sample of many sightlines.
The solid curve shows $n_p(z)$, the expected number of lines once the
ionization of foreground QSOs is included.
A background ionizing flux density of $J_{21} = 0.1$ is assumed.
 
\noindent Figure 9: shows the same information as in Figure 8, but for LB9612.
 
\noindent Figure 10: the inferred cloud radius $R$ (and $1\sigma$ confidence
intervals) from a model assuming uniformly-sized, unclustered spherical clouds,
as a function of QSO pair sightline separation $S$.
The data from this paper for the four largest $S$ pairs is shown, along with
that of Q1343+2640A/B (Paper II), Q0107-0234/0107-0235 (Dinshaw et al.~1995)
and Q0307-1931/0307-1932 (Shaver et al.~1983).
All points must sit above the solid diagonal line if they show a significant
detection, as $R > S/2$ in order for such clouds to span the sightlines.
The best linear fit to $R(S)$ is shown by the dotted line for all six pairs,
and by the dashed line once the lowest $z$ pair Q0107-0234/0107-0235 is
excluded.
 
\noindent Figure 11a: a contour plot of the probability $P_{ab}$ of a cloud
intercepting both the KP 76 and 77 sightlines if it is a cylinder of
cross-sectional radius $R$ and aspect ratio $a = l/2R$ for cylinder length $l$.
 
\noindent Figure 11b: as in Figure 11a, but for the probability $P_{abc}$ of a
cloud intercepting all three sightlines KP 76, 77 and 78.
 
\noindent Figure 12a: the measured values and 68\% confidence intervals for
$P_{ab}, P_{ac}, P_{bc}$ and $P_{abc}$ plotted on the same contours as those
shown in Figure 11.
 
\noindent Figure 12b: the measured values and 68\% confidence intervals for
$P^{\prime}_{ab}, P^{\prime}_{ac}, P^{\prime}_{bc}$ and $P_{abc}$ plotted on
the same contours as those shown in Figure 11.
 
\noindent Figure 13: the measured values and 68\% confidence intervals for
probabilities $P_{ab}, P_{ac}, P_{bc}, P^{\prime}_{ab}, P^{\prime}_{ac},
P^{\prime}_{bc}$ and $P_{abc}$ as a function of spherical cloud radius.
 
\noindent Figure 14: a cumulative histogram plot of the shear velocity
$v_{max}$ inferred for the four objects spanning the KP triplet sightlines as
a function of $v_{max}$, along with the expected plots of the same quantity
expected for sheets expanding in the Hubble flow according to three
Freidman cosmological models.

\noindent Figure 15a: a comparison of linestrengths within the Ly $\alpha$
forest of 0307-1931/32 (proper separation of 231~$h^{-1}$~kpc) showing the
difference in rest equivalent width $|W_A-W_B|$ for hits within 200~km~s$^{-1}$
versus the strength of the stronger of the two lines.
All Ly $\alpha$ line detections stronger than $3.5\sigma$ are considered.
Error bars are $\pm 1 \sigma$.
 
\noindent Figure 15b: a plot similar to Figure 15a, but for the KP triplet,
taken one QSO pair at a time.
Due to the density of points, no error bars are shown, but errors are much
smaller than in Figure 15a, typically the size of the round symbols.
The three-legged crosses mark lines involved in objects which span all three
sightlines.
 
\noindent Figure 16: the distribution of anticoincident lines (all lines
stronger than $3.5\sigma$ detections, but missing a neighbor stronger than
$W_o =0.3$\AA within 200~km~s$^{-1}$ in the adjacent sightline).
In the case of the KP triplet, each sightline pair is considered individually,
Rest equivalent width has been converted to $N_{HI}$ by assuming thermal widths
of $b=30$~km~s$^{-1}$, and the distributions are normalized to be equal at
$log (N_{HI}/cm^{-2})=13.5$.

\noindent Figure 17: the C~IV$\lambda$1548 absorber two-point velocity
cross-correlation function for three different samples, plotted as a histogram
of the number of pairs in 200~km~s$^{-1}$ bins versus velocity difference
$\Delta v$.
The dotted bars indicate only those pairs composed of ``definite'' or
``probable'' C~IV$\lambda$1548 absorbers among the three KP triplet sightlines,
while the solid bars indicate pairs including ``possible'' KP triplet
absorbers.
The dashed bars include pairs from Q1517+2357/1517+2356 and
Q0307-1931/0307-1932 (Shaver et al.~1983).
 
\newpage



\begin{deluxetable}{ccccccc}
\tablecaption{FOREGROUND QSO PROXIMITY EFFECT: PREDICTED VERSUS OBSERVED
DENSITIES OF $W_o > 0.1$\AA\ LY $\alpha$ LINES}
\tablewidth{0pt}
\tablehead{
\colhead{} &
\colhead{} &
\colhead{Number} &
\colhead{Proximity} &
\colhead{Null Model} &
\colhead{Observed} &
\colhead{} \\
\colhead{} &
\colhead{Redshift} &
\colhead{of Lines,} &
\colhead{Prediction} &
\colhead{Prediction} &
\colhead{Value} &
\colhead{Variance} \\
\colhead{QSOs} &
\colhead{Range} &
\colhead{$n$} &
\colhead{($\int~w n_p dz$)} &
\colhead{($\int~w n_\gamma dz$)} &
\colhead{($\sum_{i=1}^n ~ w_i$)} &
\colhead{($\sum_{i=1}^n ~ w_i^2$)}
} 
\startdata
KP 76 & 2.175-2.191 &  2 & 0.150 &  0.163 &  0.166 & 0.014 \nl
KP 77 & 2.173-2.193 & 14 & 1.562 &  2.532 &  3.657 & 1.461 \nl
      & 2.425-2.485 &    &       &        &        &       \nl
KP 78 & 2.173-2.193 & 23 & 4.038 &  7.708 &  7.561 & 3.679 \nl
      & 2.435-2.574 &    &       &        &        &       \nl
KP triplet&(above ranges)&39&5.750&10.403 & 11.384 & 5.154 \nl
combined&           &    &       &        &        &       \nl
      &    &             &       &        &        &       \nl
LB 9612& 1.815-1.857&  2 & 0.609 &  1.047 &  0.286 & 0.047 \nl
      &             &    &       &        &        &       \nl
All 4 QSOs&(above ranges)&41&6.359&11.450 & 11.670 & 5.201 \nl
\enddata
\end{deluxetable}

\begin{deluxetable}{lcccccccc}
\tablewidth{0pt}
\tablecaption{Ly$\alpha$ CLOUD RADIUS ESTIMATES FROM QSO PAIRS}
\tablehead{
\colhead{} &
\colhead{Angular} &
\colhead{Ly$\alpha$ $z$} &
\colhead{$S$, Proper} &
\colhead{} &
\colhead{} &
\colhead{95\% Confidence} &
\colhead{Median} \\
\colhead{QSO Pair} &
\colhead{Separation} &
\colhead{Range} &
\colhead{Separation} &
\colhead{$N_h$} &
\colhead{$N_m$} &
\colhead{Interval in $R$} &
\colhead{Radius, $R$} \\
\colhead{} &
\colhead{($^{\prime\prime}$)} &
\colhead{} &
\colhead{($h^{-1} kpc$)} &
\colhead{} &
\colhead{} &
\colhead{($h^{-1} kpc$)} &
\colhead{($h^{-1} kpc$)}
}
\startdata
1343+2640A/B  & 9.5 & 1.756-2.035 & 39-40   & ~7 & ~1 & ~77-841 & 237 \nl
1025-0045A/B  & 36  & 0.830-1.438 & 149-154 & ~1 & ~5 & ~95-306 & 148 \nl
0307-1931/32  & 56  & 1.690-2.122 & 226-236 & ~4 & 12 & 165-434 & 246 \nl
0107-0234/35  & 86  & 0.481-0.952 & 301-364 & ~4 & ~6 & 286-918 & 501 \nl
1517+2356/57  & 102 & 1.390-1.830 & 425-438 & ~3 & 28 & 226-429 & 283 \nl
1623+2651A/B  & 127 & 2.025-2.467 & 493-522 & ~6 & 32 & 264-520 & 340 \nl
1623+2651A/53 & 147 & 1.958-2.467 & 571-604 & ~8 & 26 & 323-757 & 461 \nl
1623+2653/51B & 177 & 2.025-2.526 & 683-721 & ~5 & 36 & 357-675 & 442 \nl
\enddata
\end{deluxetable}

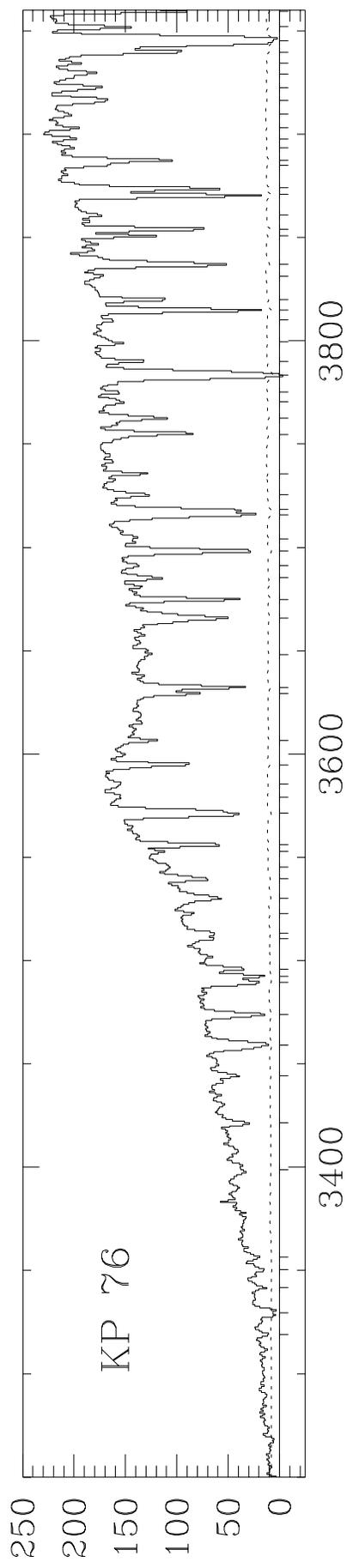
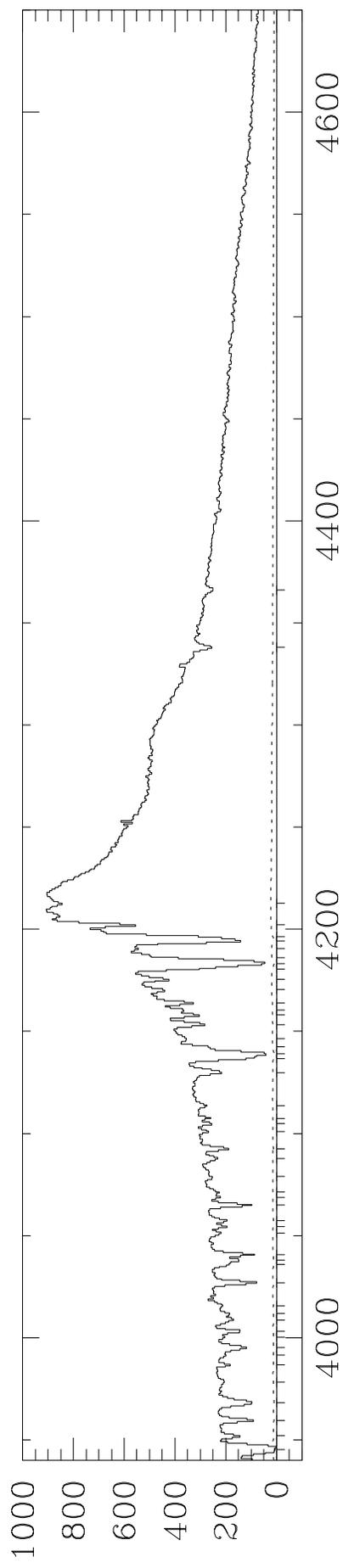
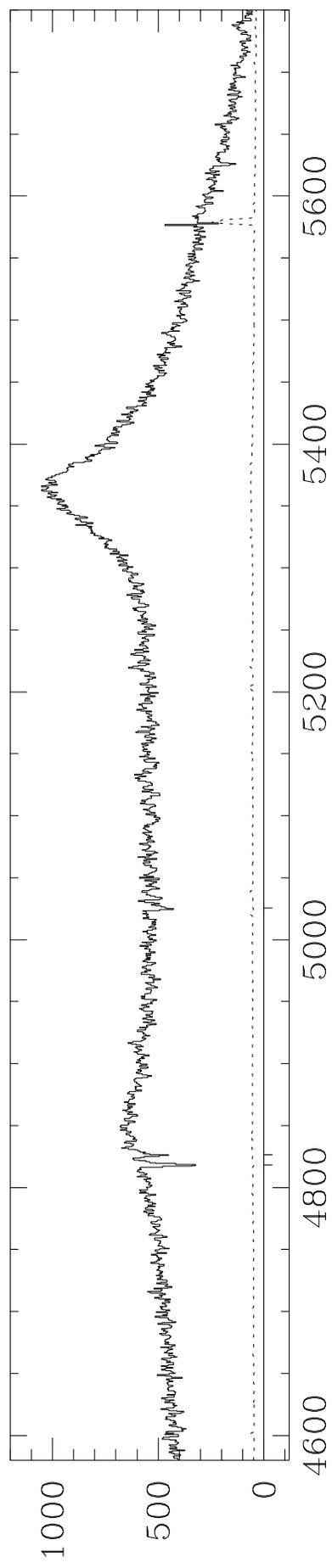
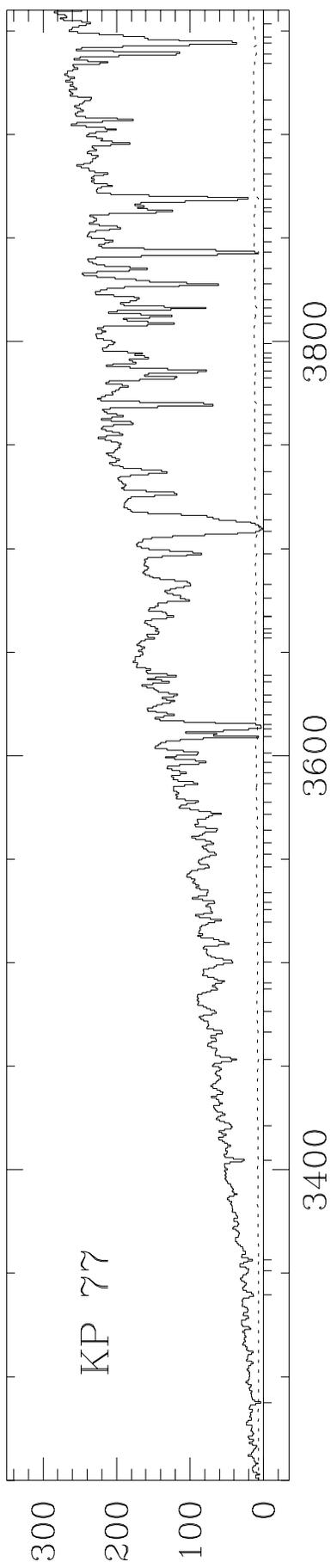
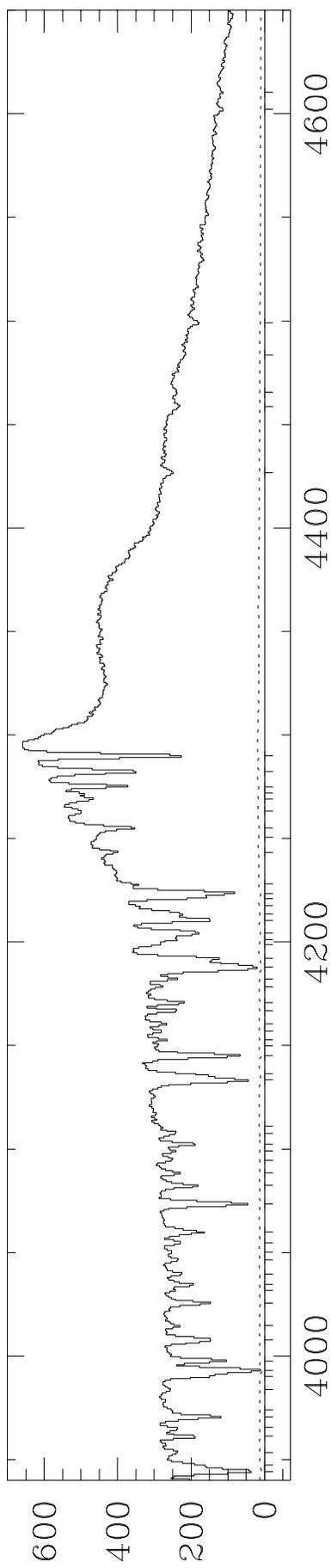
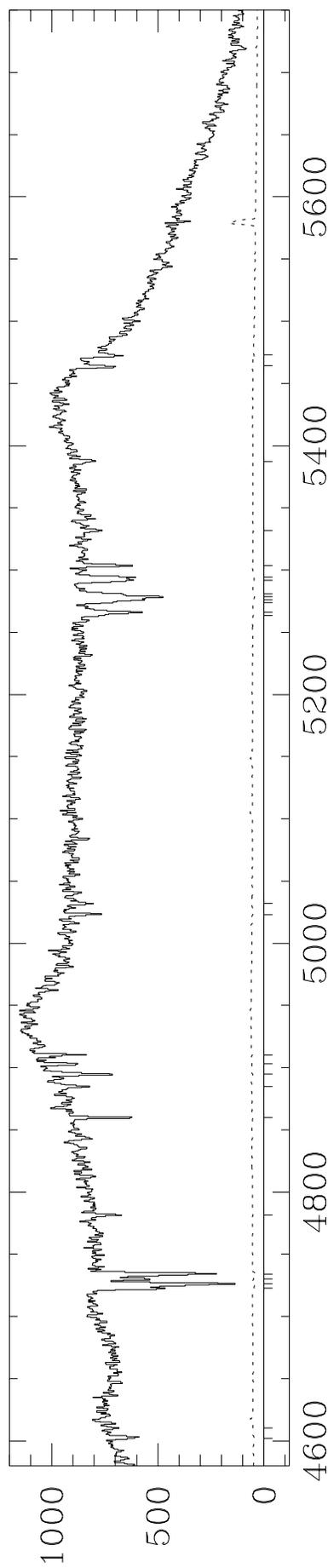
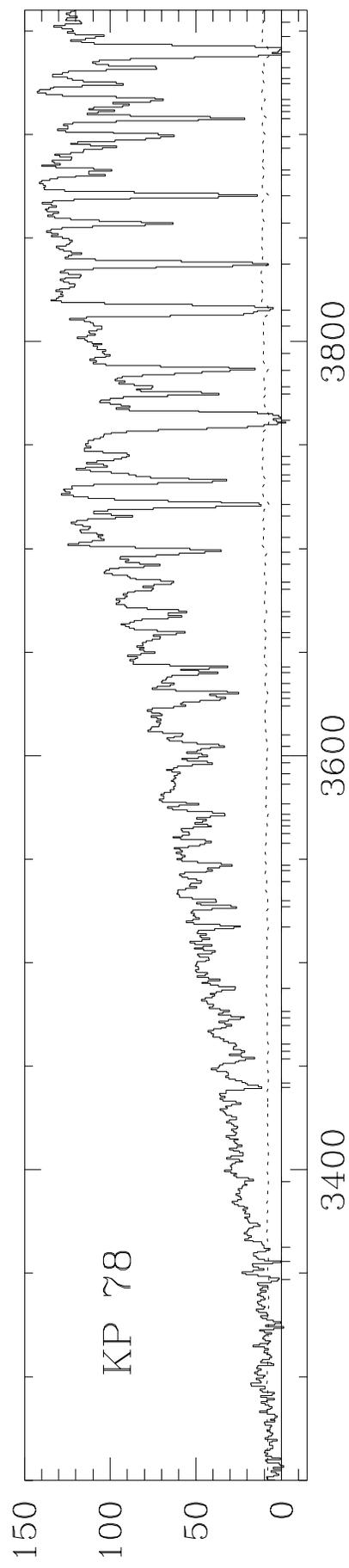
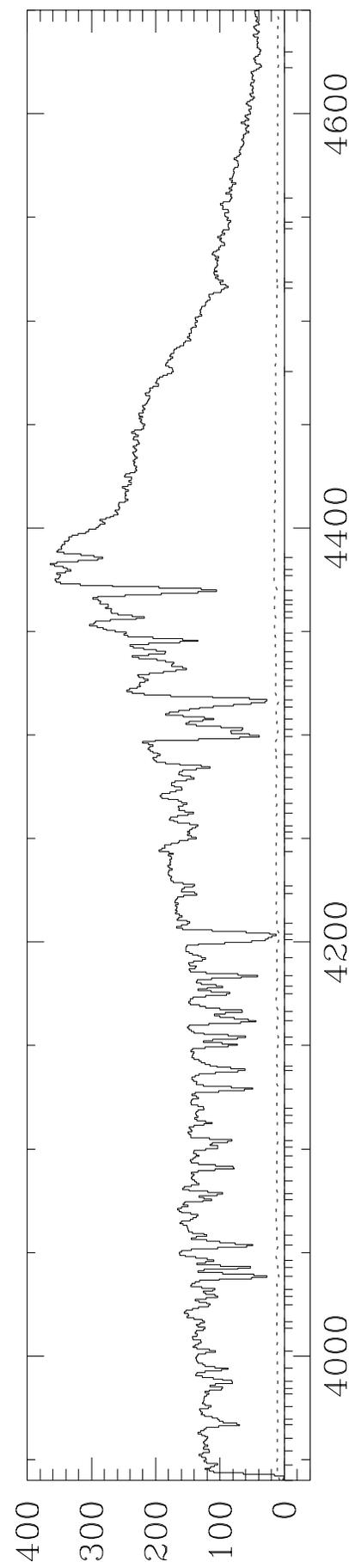
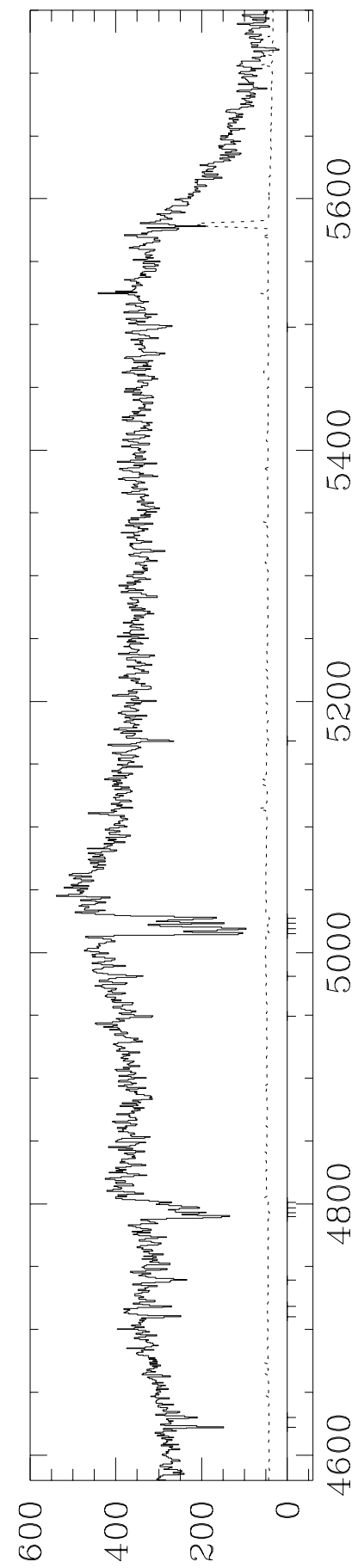
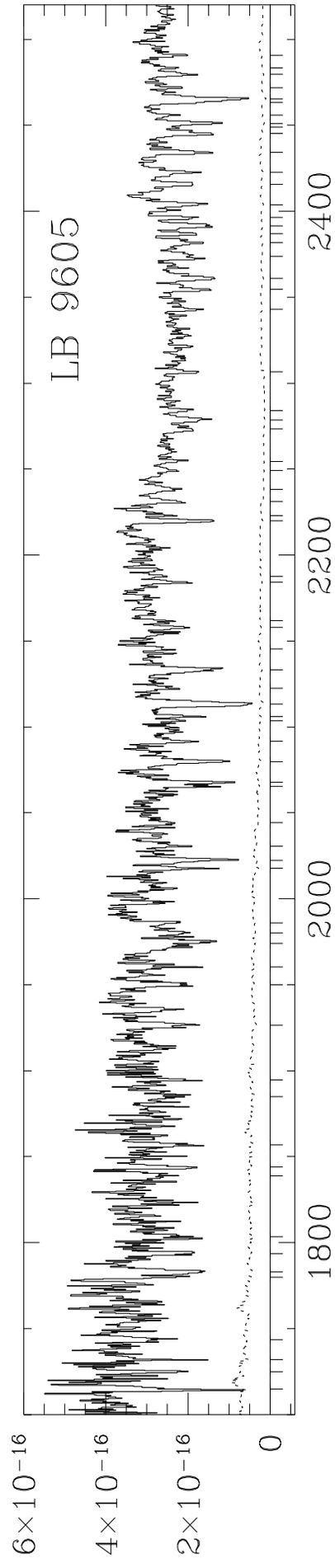
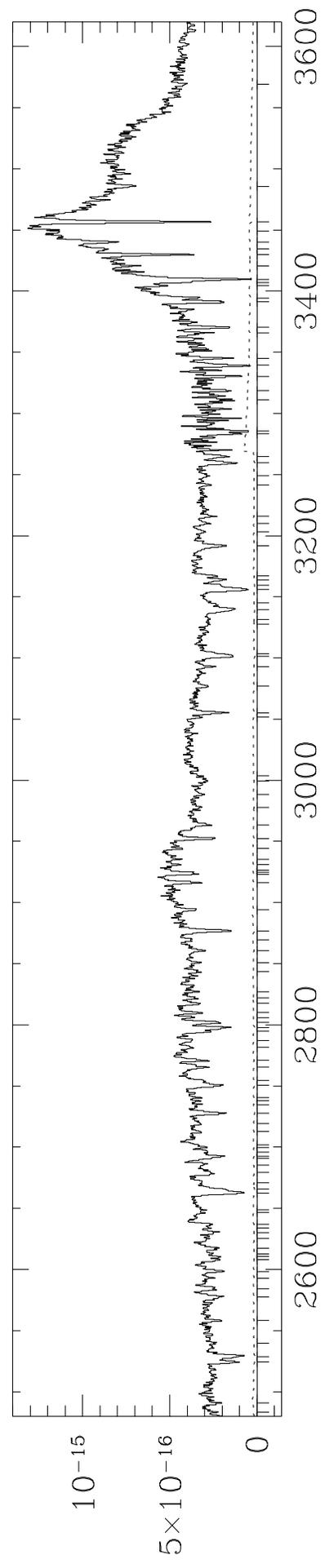
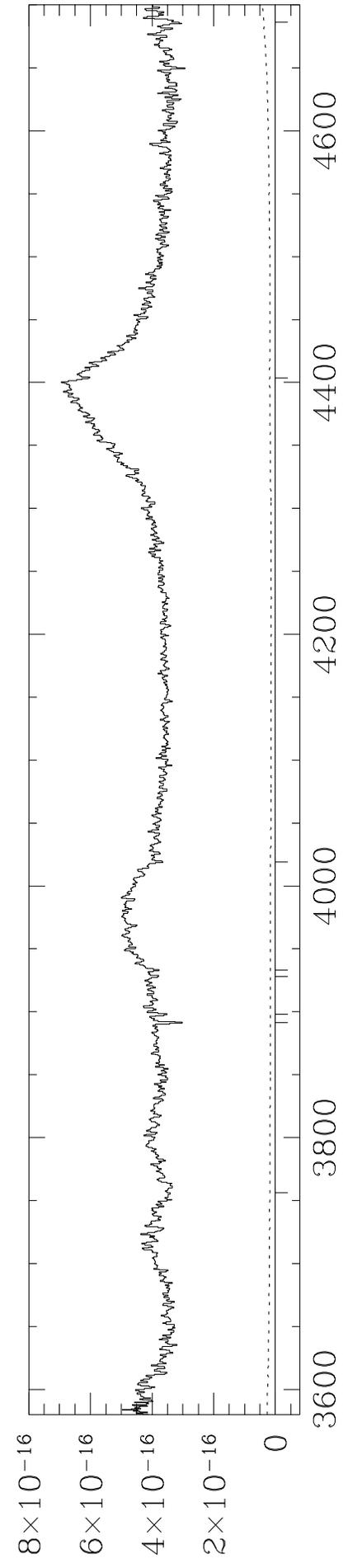
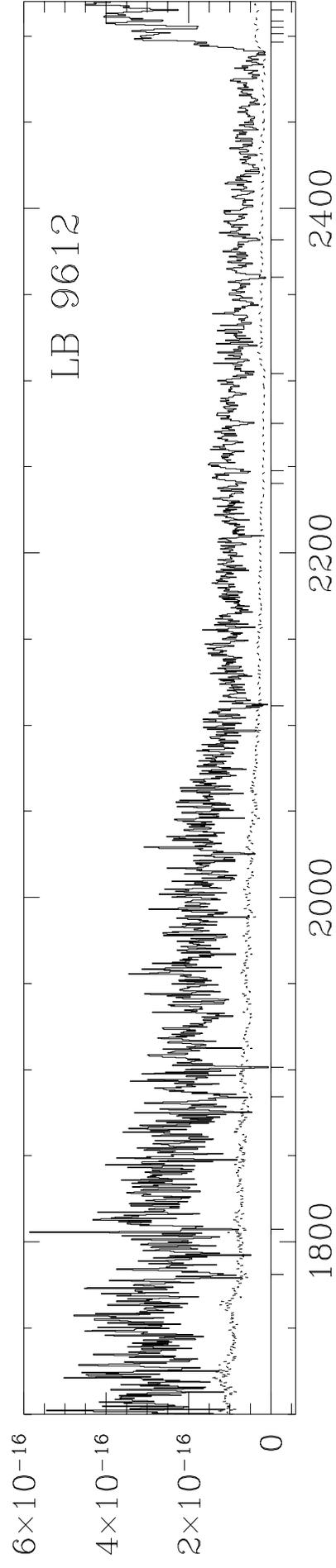
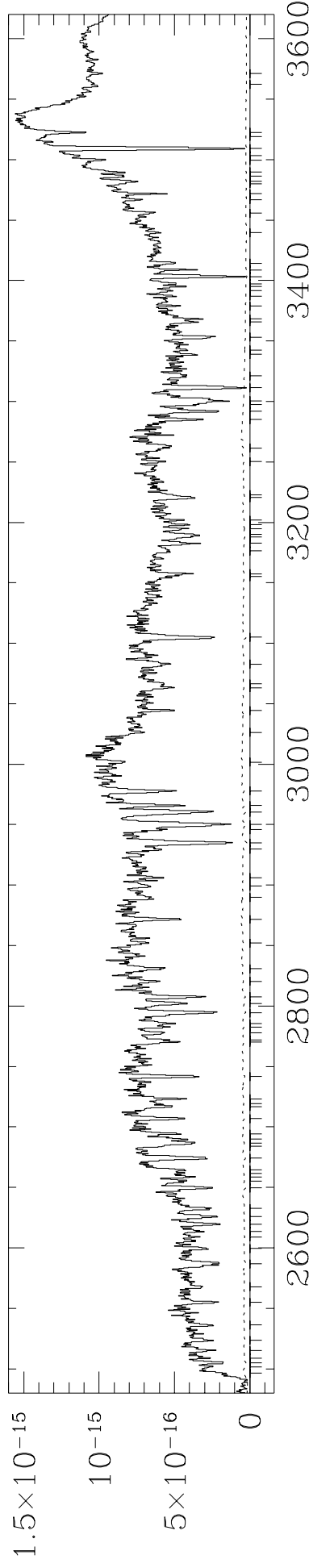
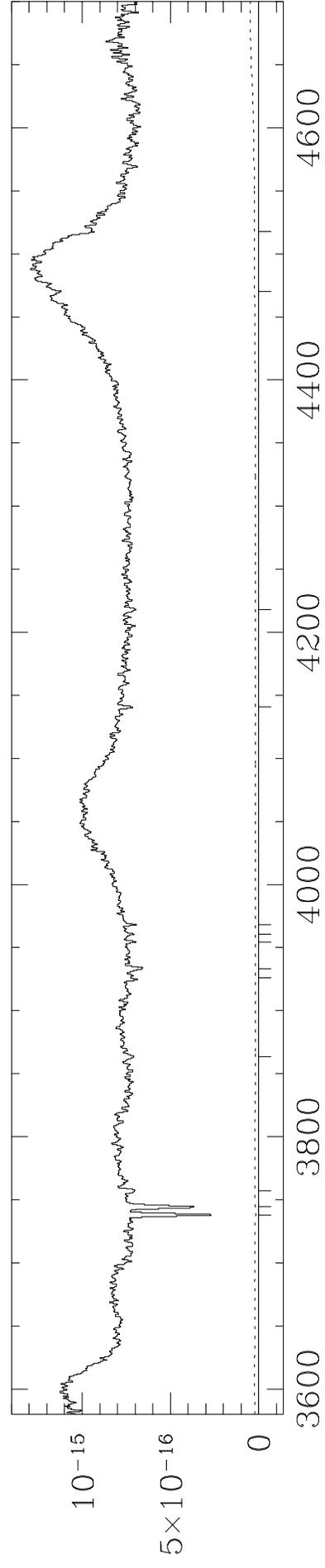
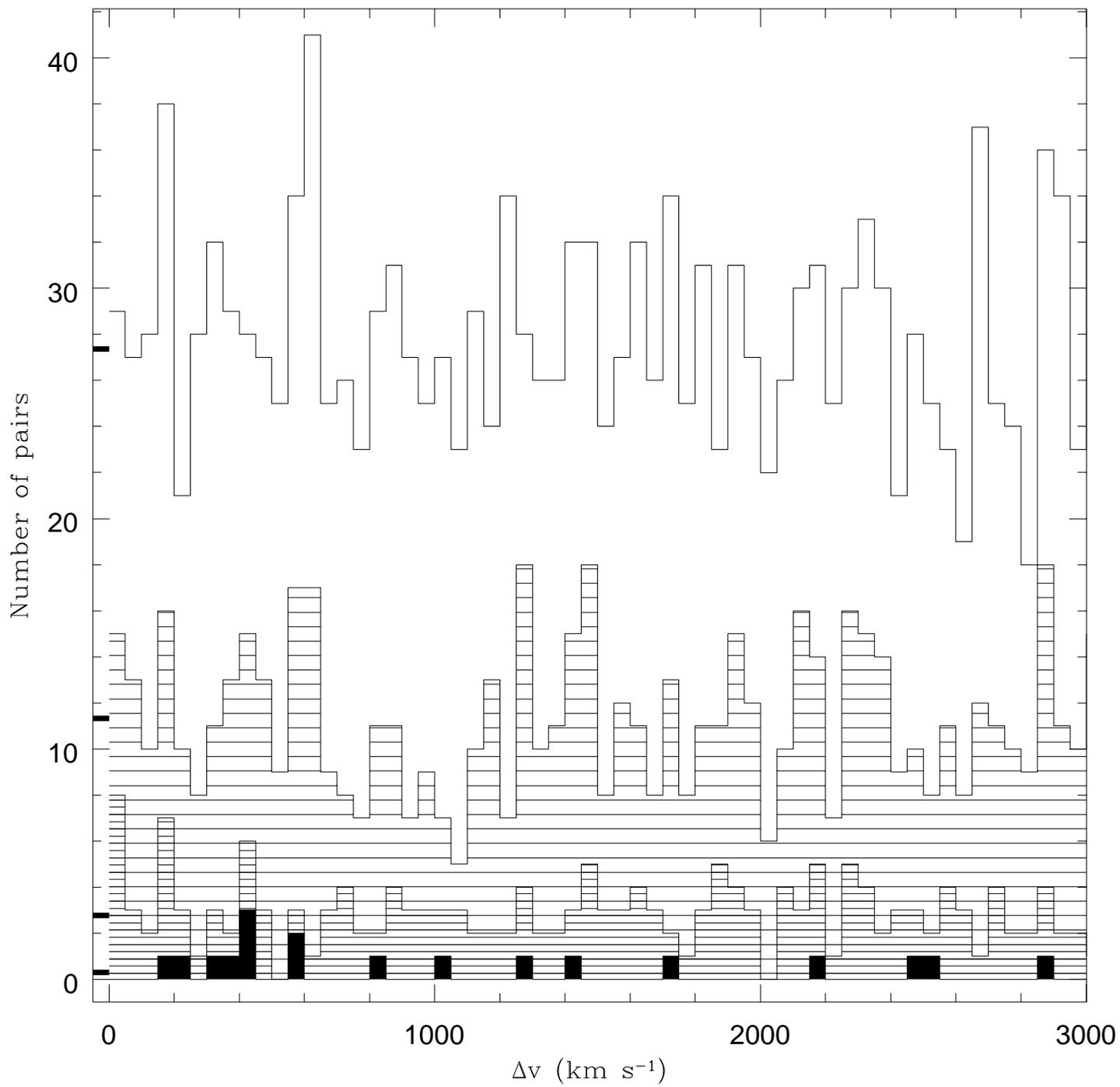
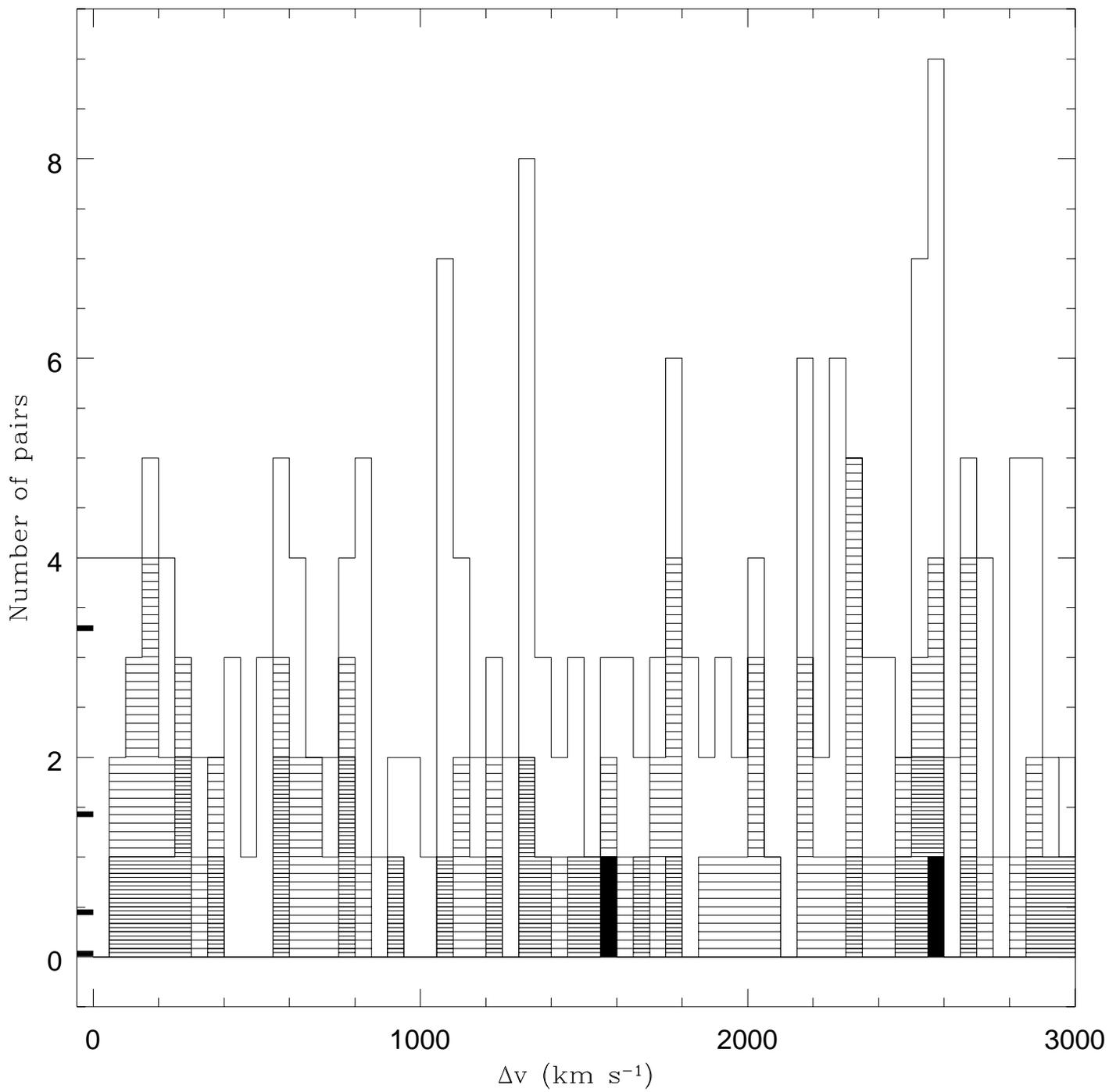
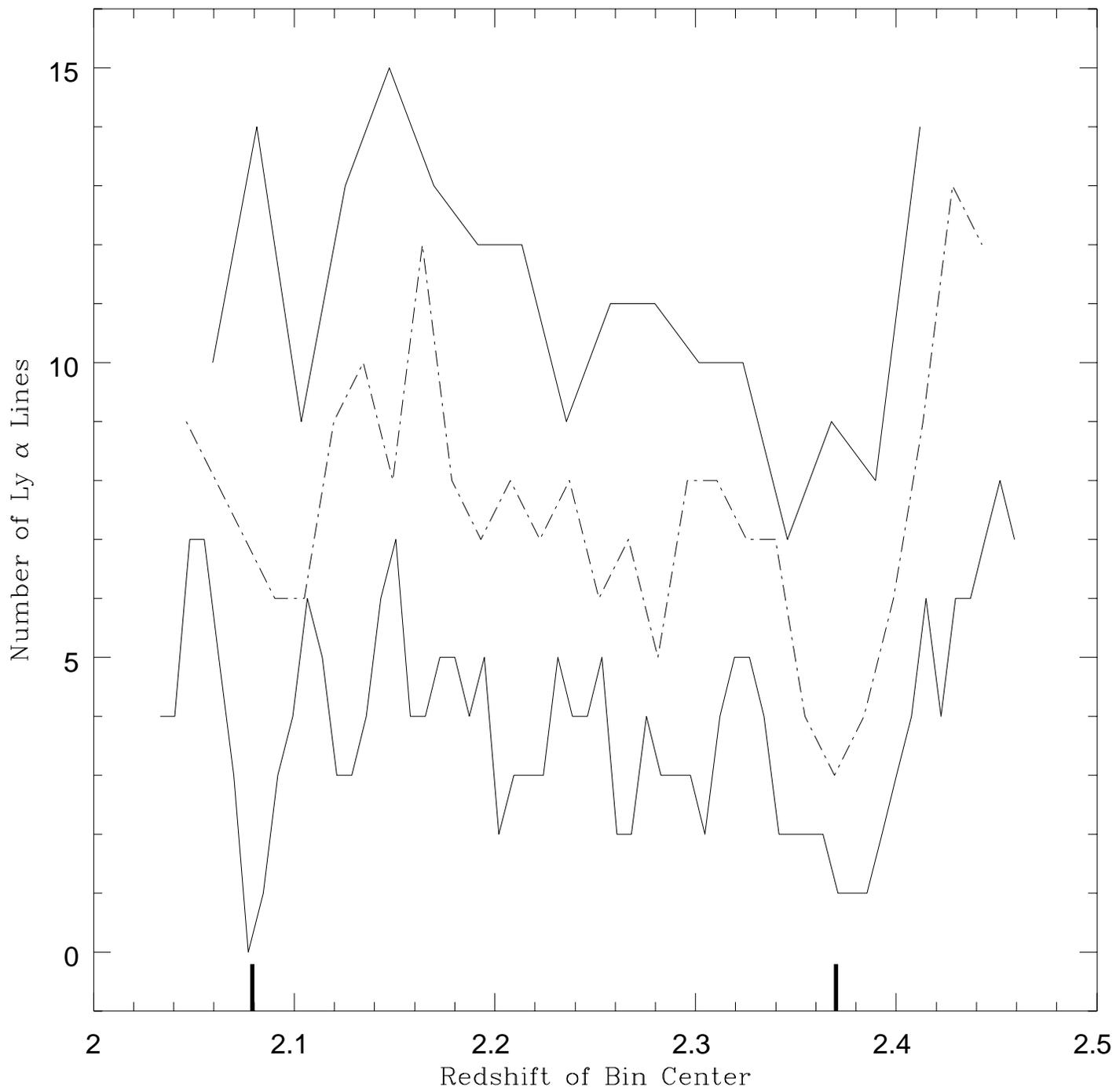
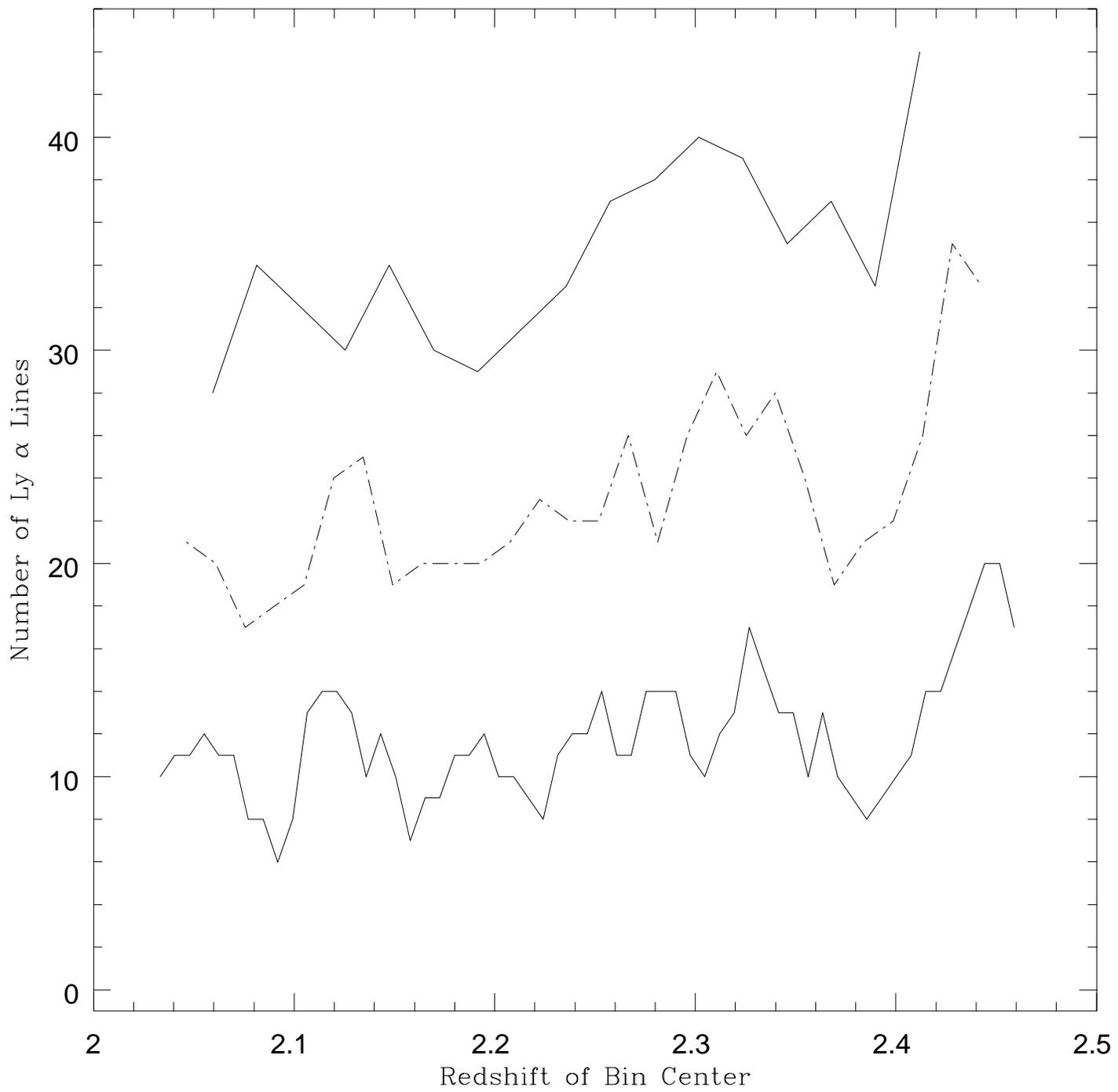
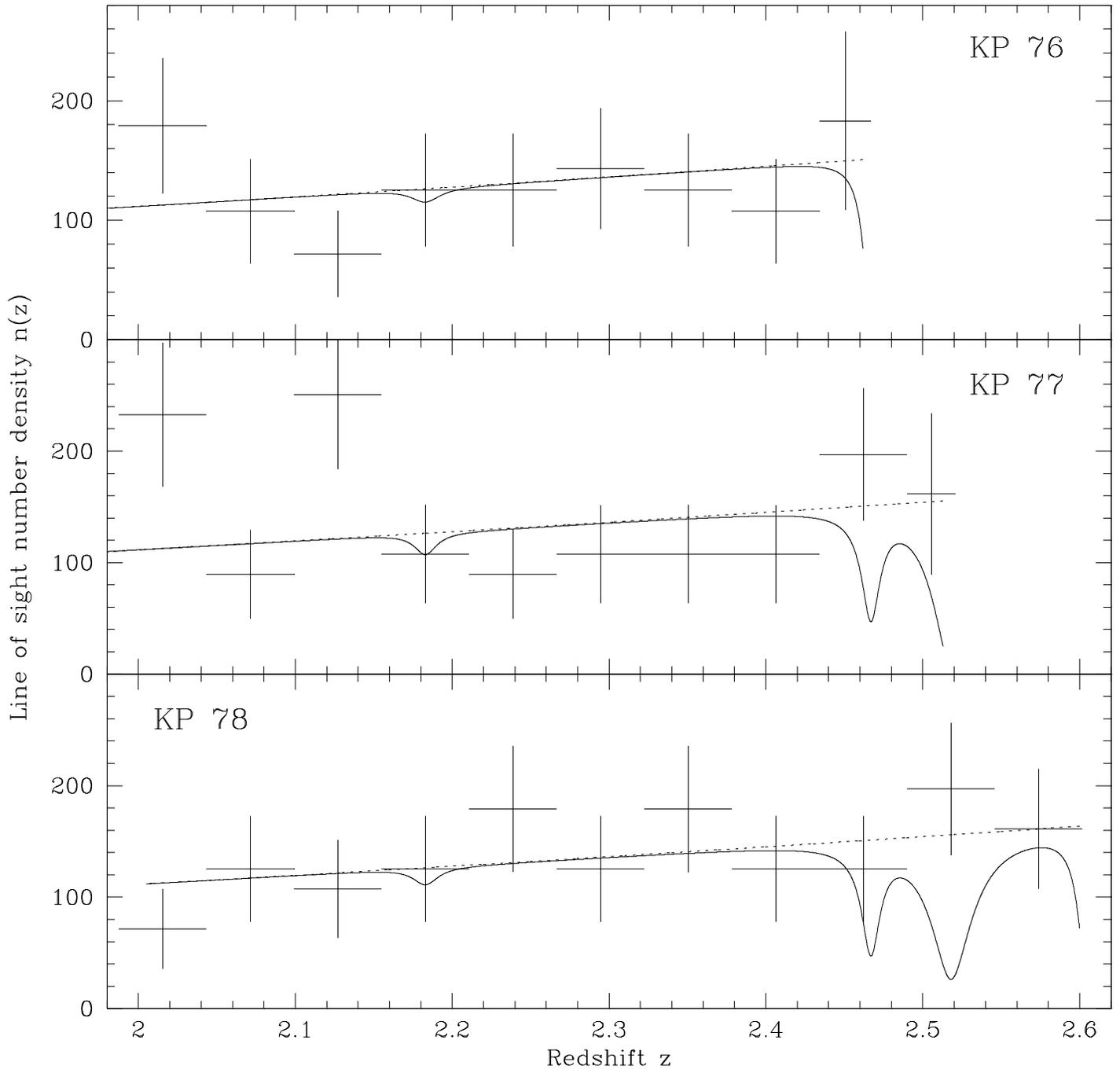
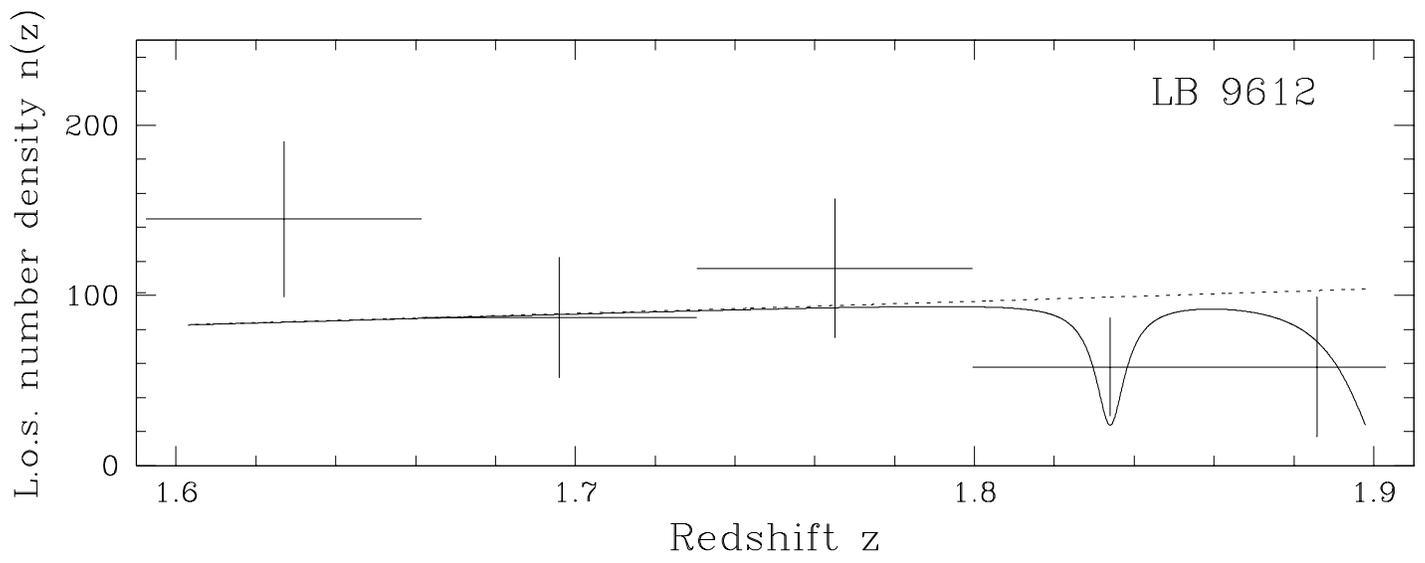
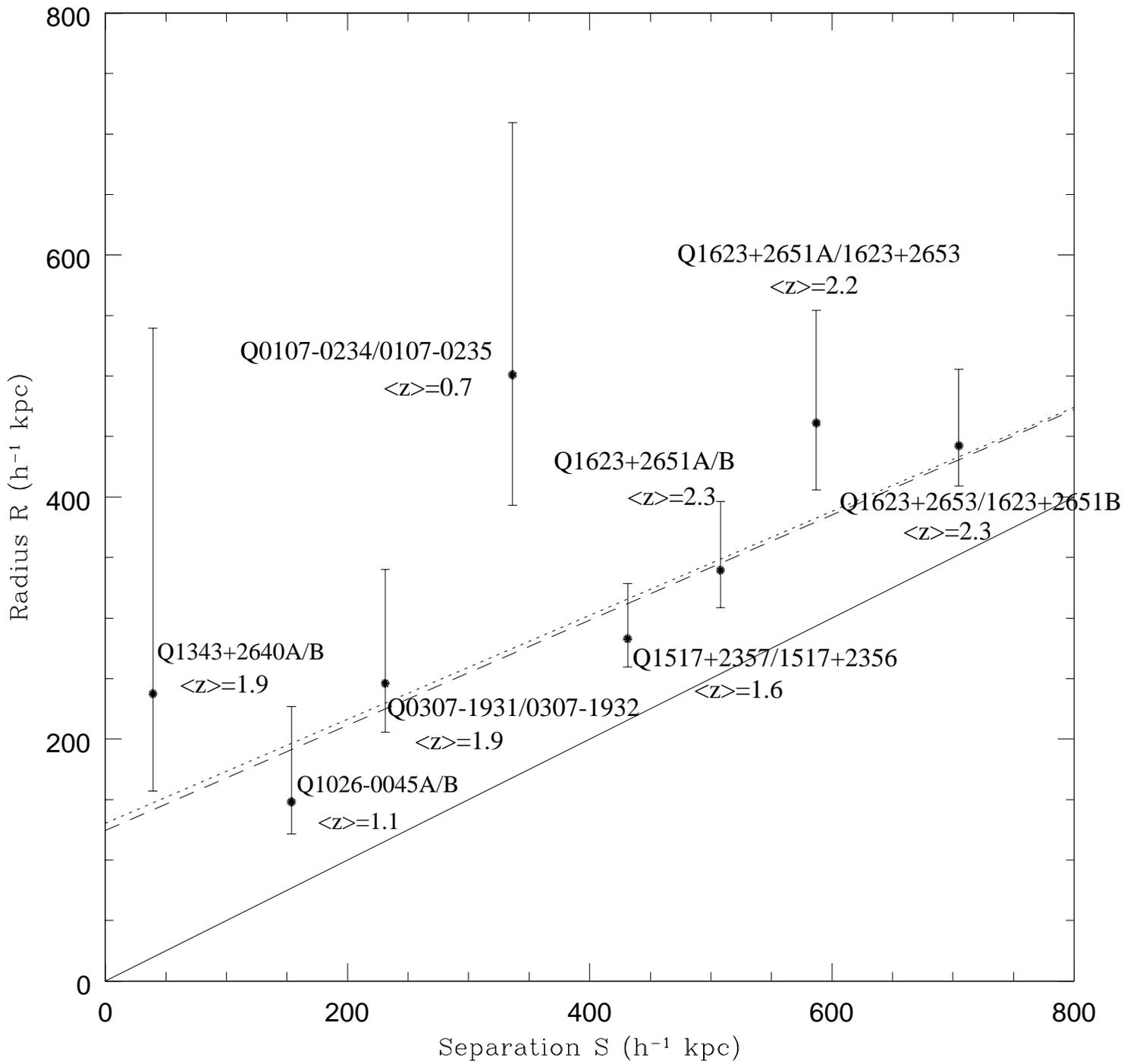
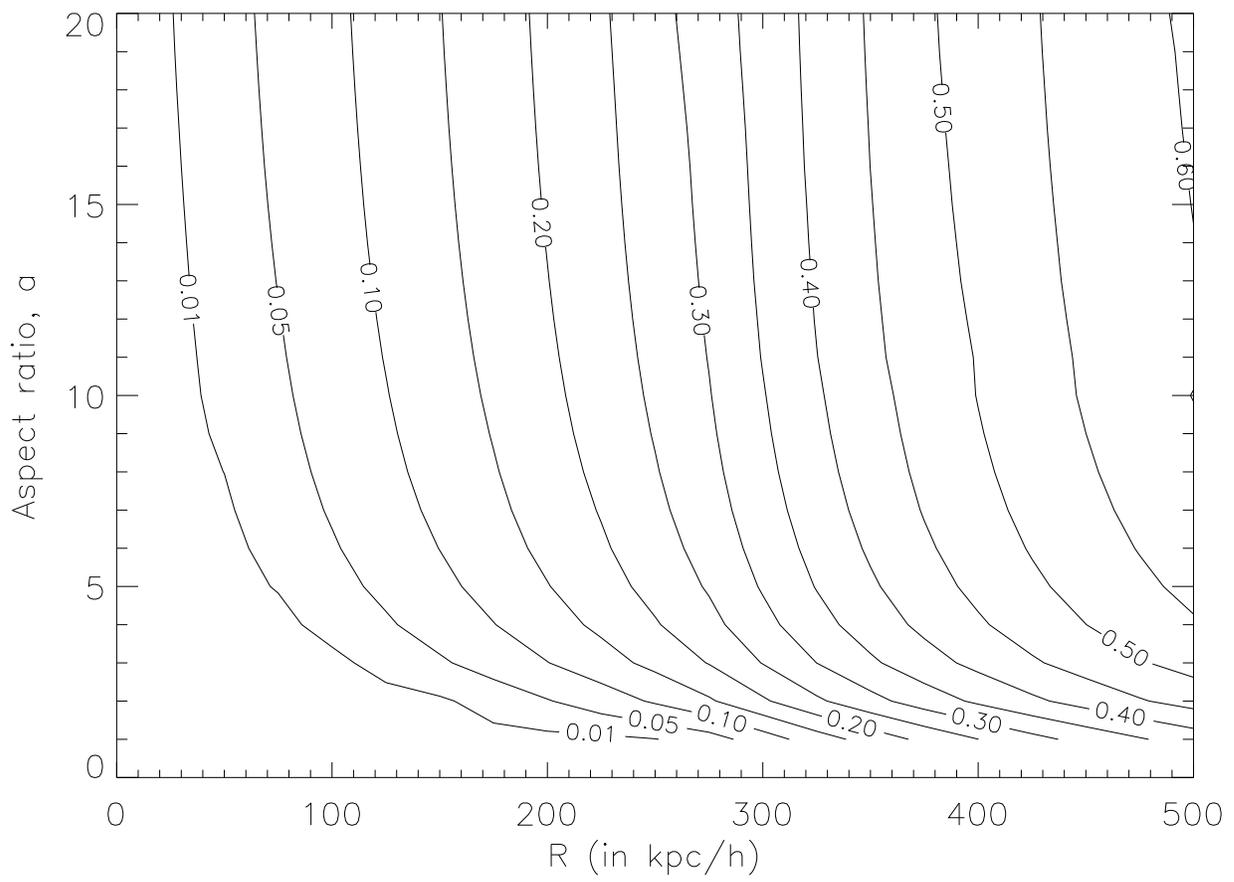
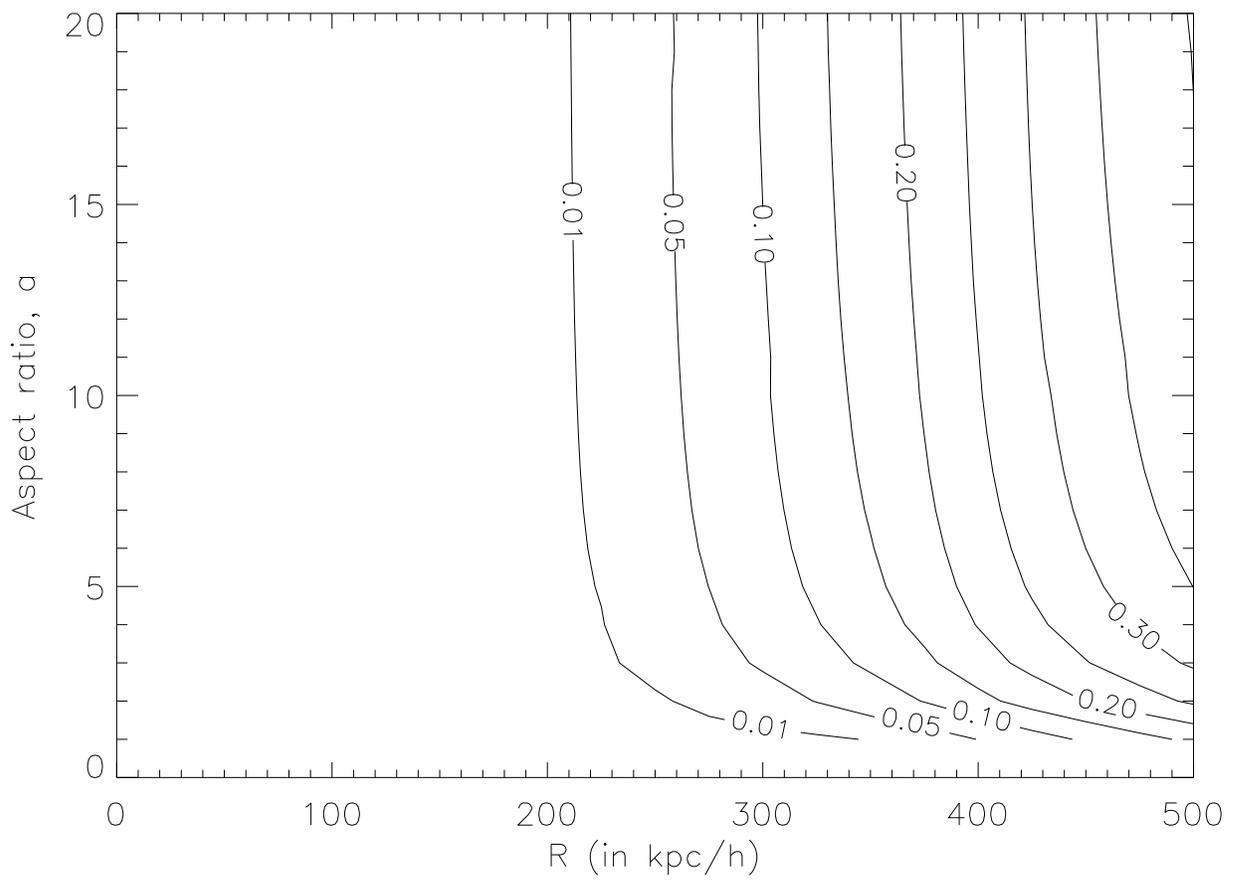
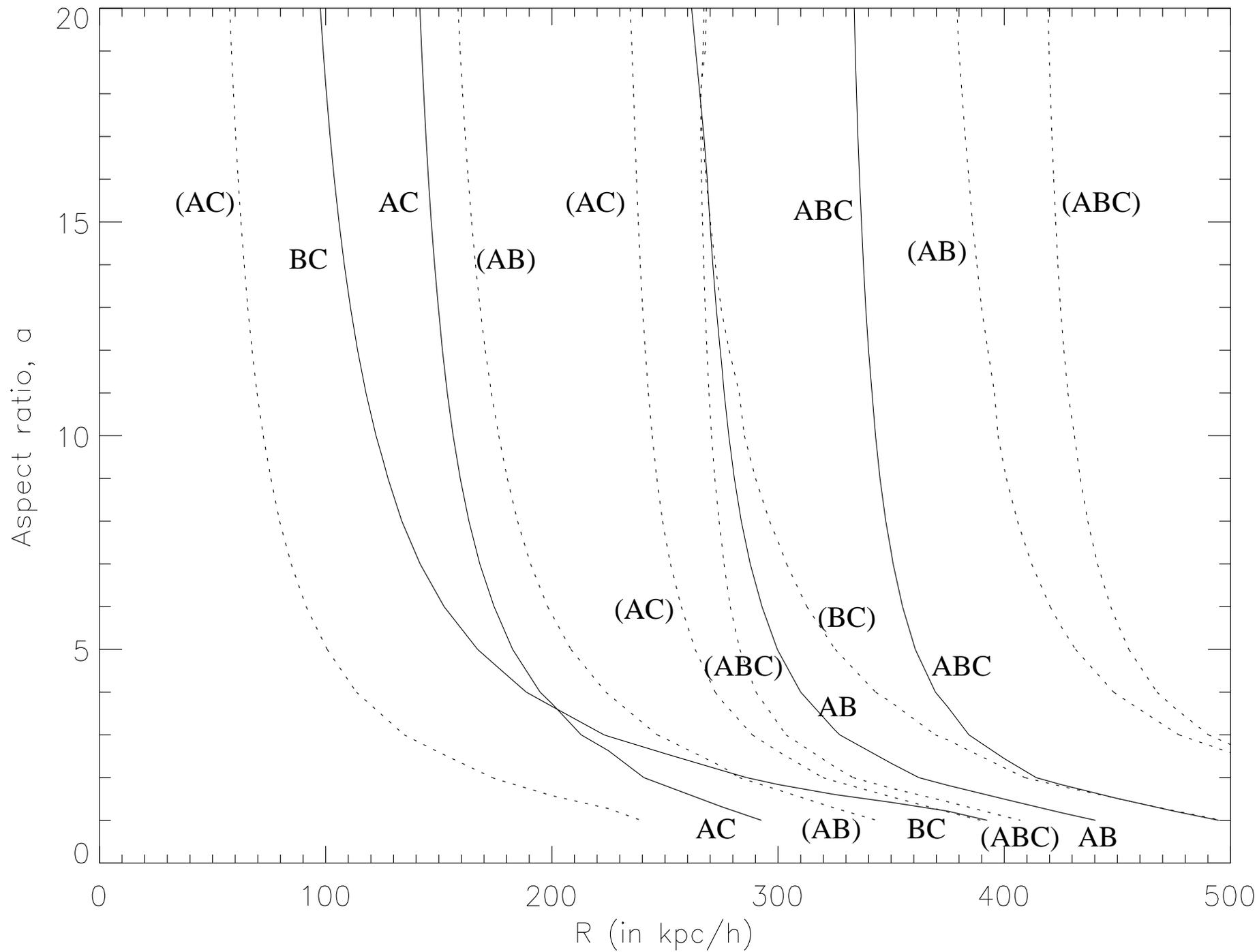
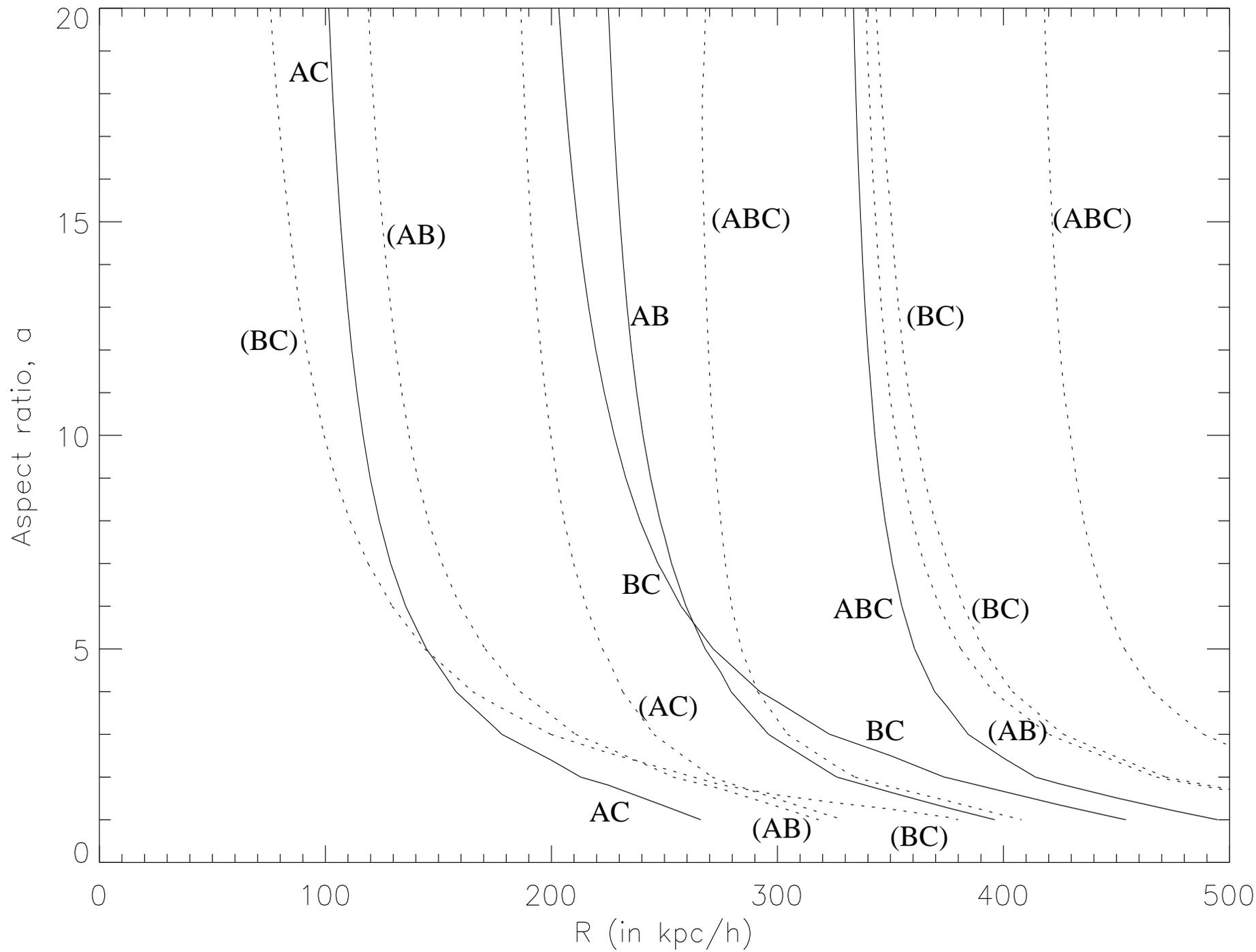
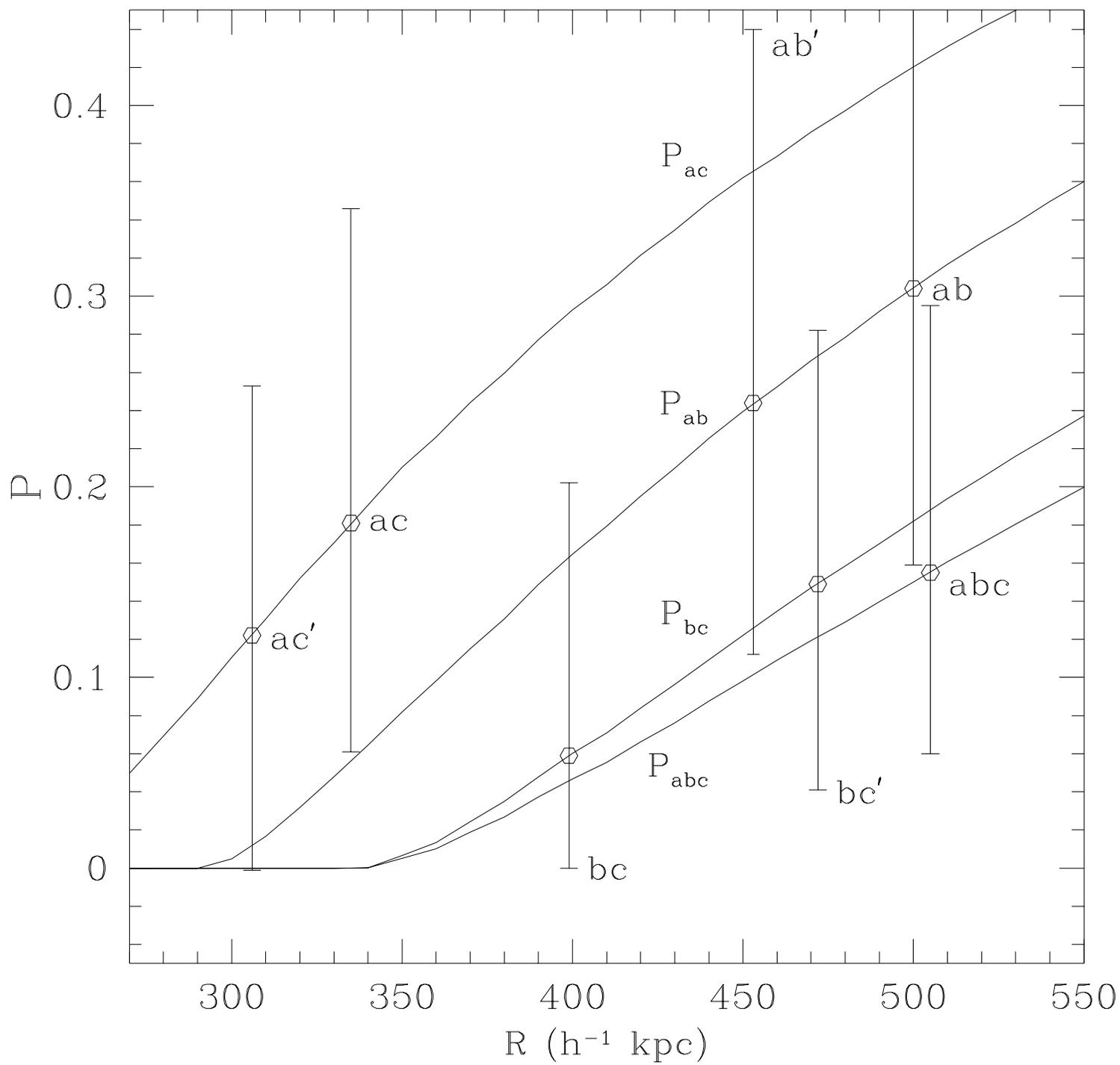
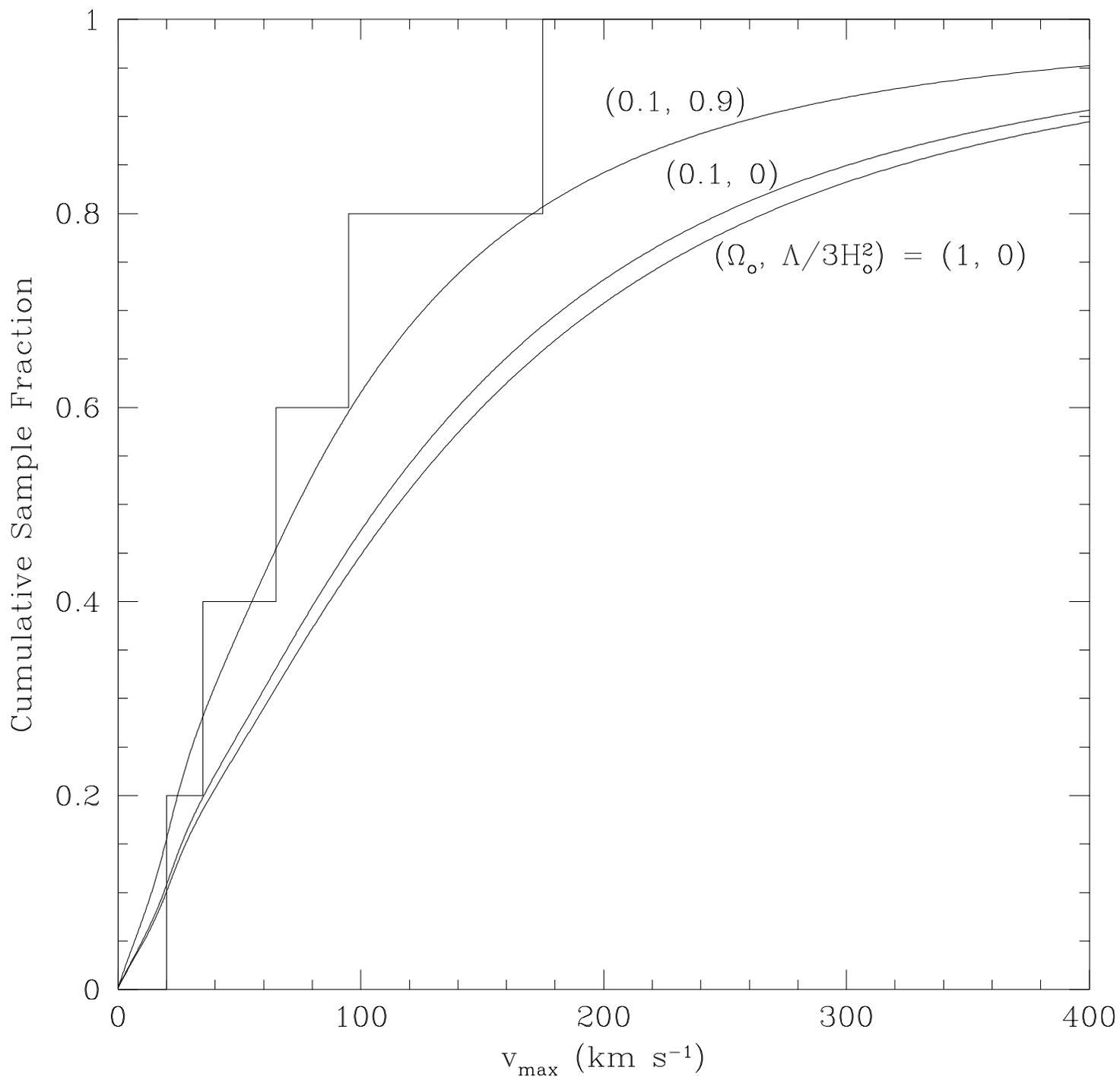
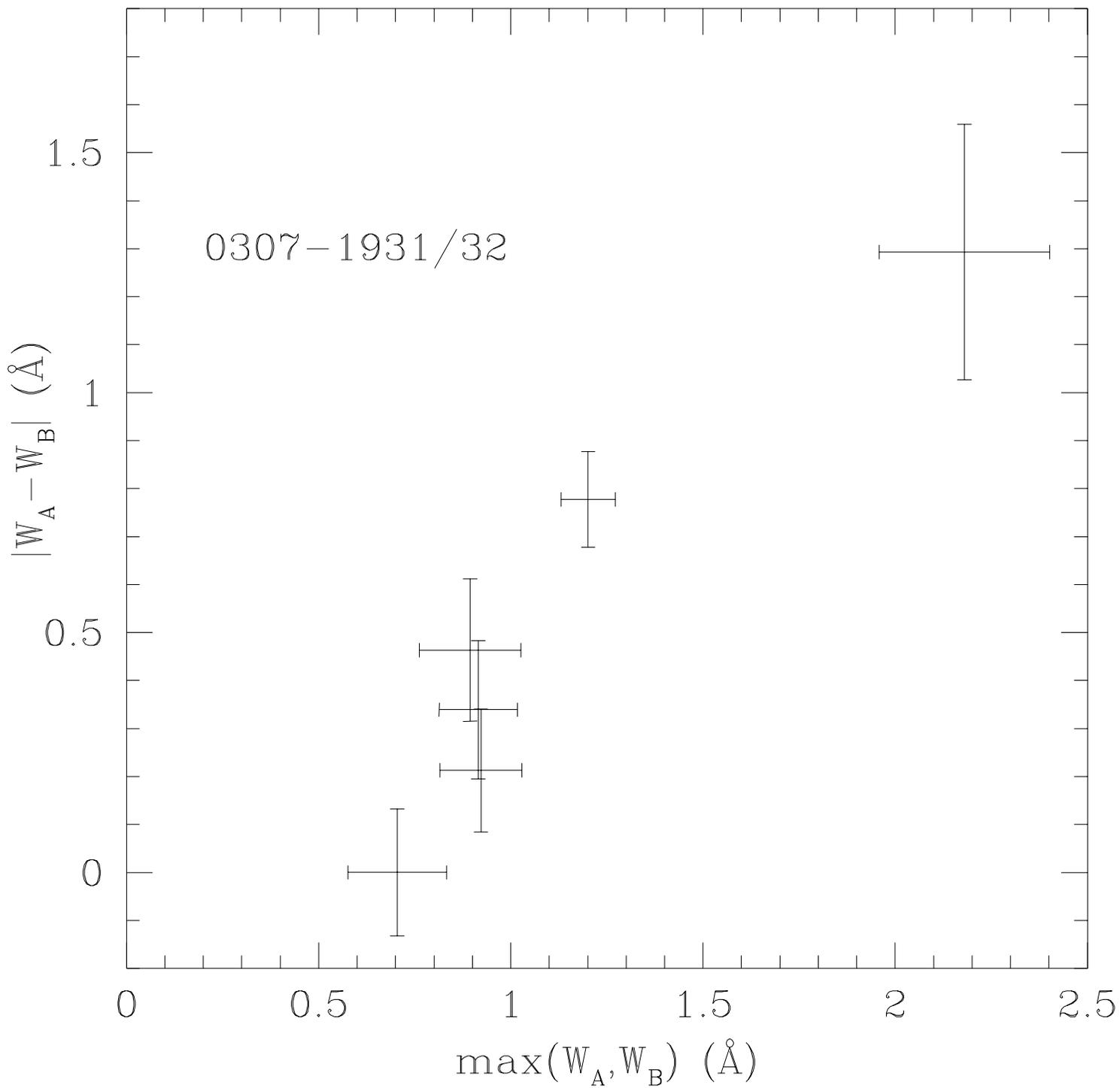
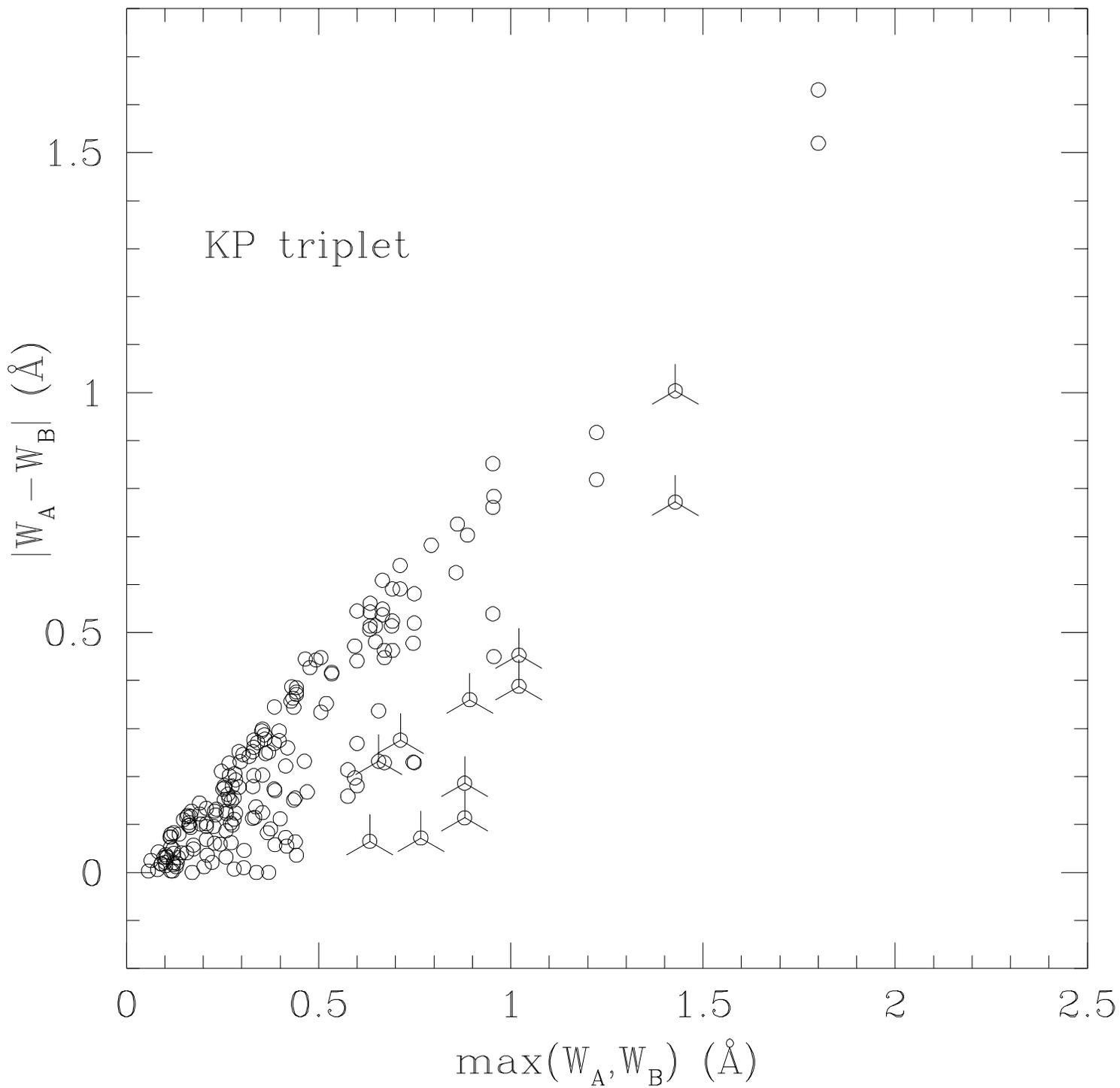
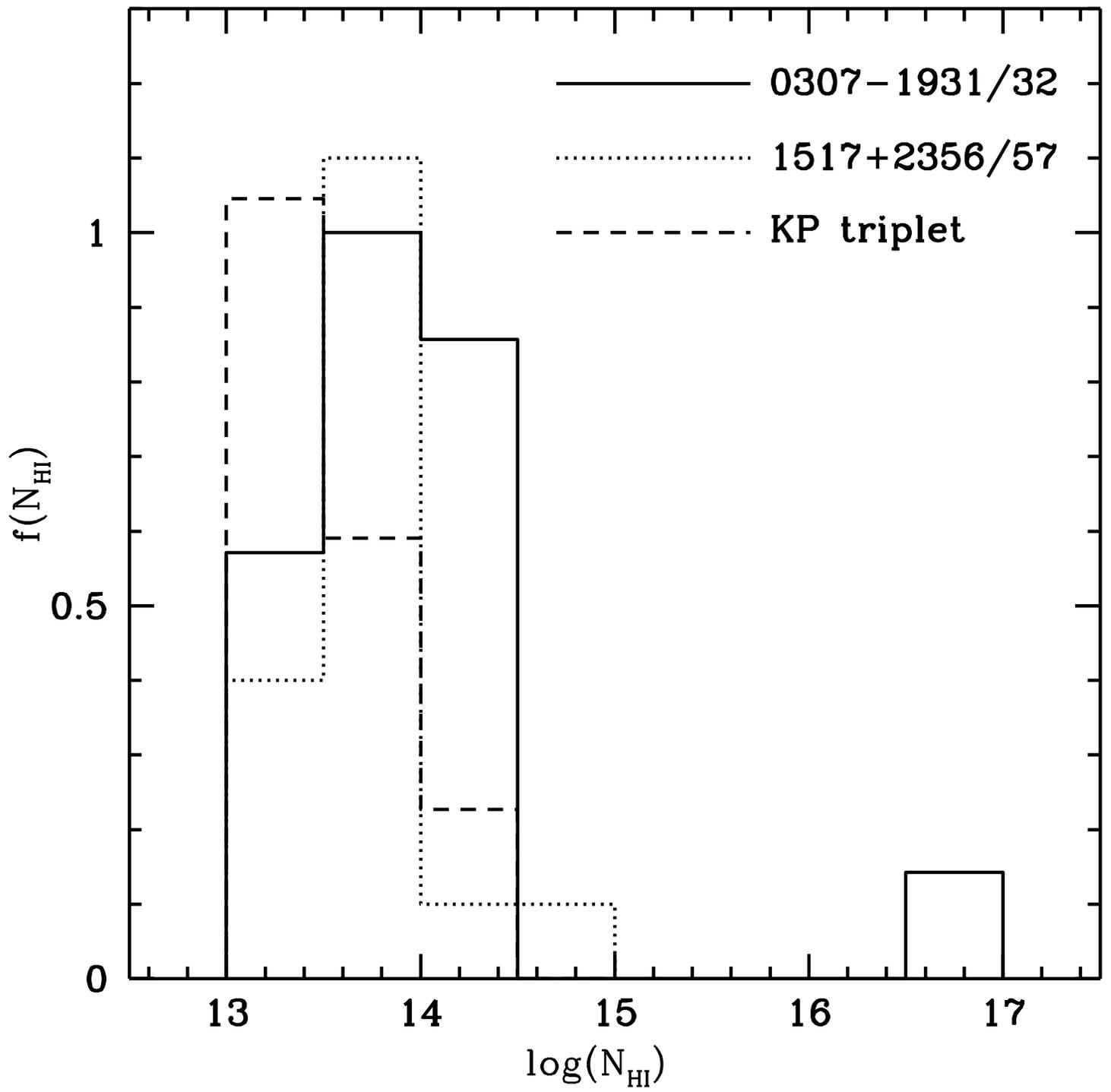
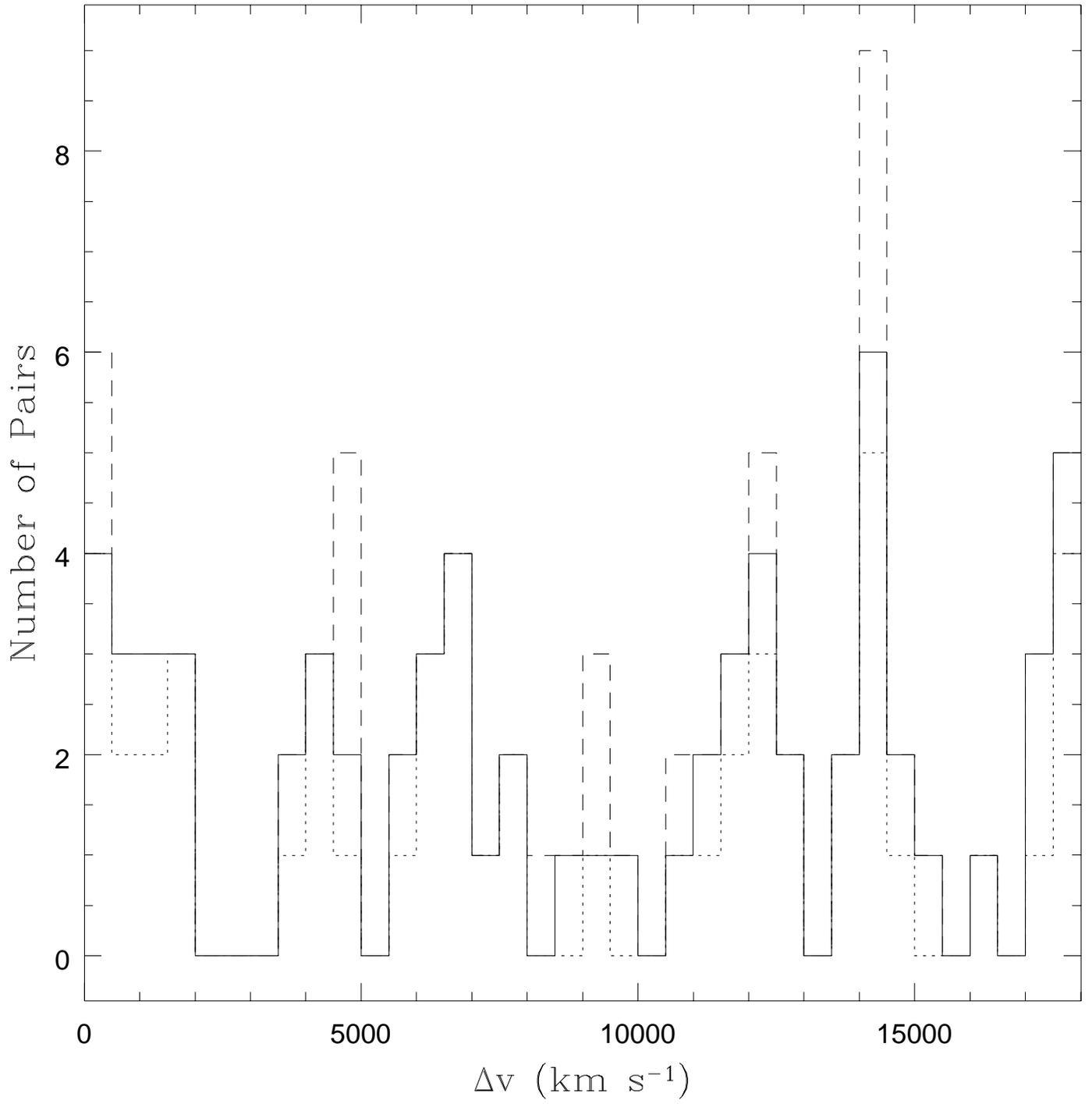
\end{document}